\documentclass[12pt]{iopart}
\usepackage{cite}
\usepackage{graphicx}
\usepackage{iopams}

\begin{document}

\title[Combined tripod and $\Lambda$ atom-light coupling
scheme]{Electromagnetically induced transparency and nonlinear pulse
  propagation in a combined tripod and $\Lambda$ atom-light coupling scheme}

\author{H. R. Hamedi, J. Ruseckas and G. Juzeli\={u}nas}

\address{Institute of Theoretical Physics and Astronomy, Vilnius University,
Saul\.{e}tekio 3, Vilnius LT-10222, Lithuania}

\ead{hamid.hamedi@tfai.vu.lt}

\begin{abstract}
We consider propagation of a probe pulse in an atomic medium characterized
by a combined tripod and Lambda ($\Lambda$) atom-light coupling scheme.
The scheme involves three atomic ground states coupled to two excited
states by five light fields. It is demonstrated that dark states can
be formed for such an atom-light coupling. This is essential for formation
of the electromagnetically induced transparency (EIT) and slow light.
In the limiting cases the scheme reduces to conventional $\Lambda$-
or $N$-type atom-light couplings providing the EIT or absorption,
respectively. Thus the atomic system can experience a transition from
the EIT to the absorption by changing the amplitudes or phases of
control lasers.  Subsequently the scheme is employed to analyze the
nonlinear pulse propagation using the coupled Maxwell-Bloch equations.
It is shown that generation of stable slow light optical solitons
is possible in such a five-level combined tripod and $\Lambda$ atomic
system. 
\end{abstract}
\pacs{42.50.Gy; 42.65.Tg; 42.81.Dp}

\submitto{\jpb}

\maketitle

\section{Introduction}

Electromagnetically induced transparency (EIT)
\cite{Arimondo-96,Harris-PT-1997,Lukin2003,Fleischhauer2005,Wu2005EIT,Fleischhauer2016}
plays an important role in controlling the propagation of light puses in
resonant media. Due to the EIT a weak probe beam of light tuned to an atomic
resonance can propagate slowly and is almost lossless when the medium is driven
by one or several control beams of light with a higher intensity
\cite{Arimondo-96,Harris-PT-1997,Lukin2003,Fleischhauer2005,Wu2005EIT,Fleischhauer2016}.
The EIT is formed because the control and probe beams drive the atoms to their
dark states representing a special superposition of the atomic ground states
immune to the atom-light coupling. The absorption is suppressed due to a
quantum mechanical interference between different excitation pathways of atomic
energy levels leading to the EIT. The EIT has various important applications in
quantum and nonlinear optics, such as slow and stored light
\cite{Hau-99,Fleischhauer2000,Liu-01,Phill2001,Fleischhauer2002,Juzeliunas-02},
stationary light \cite{Bajcsy2003,Yu2009}, multiwave mixing
\cite{Wu2003,Zhang2008,Zhang2009}, optical solitons
\cite{wu2004,Huang2005,Li2010,Si2010,Yang2011,Chen2014}, optical bistability
\cite{Joshi2003,li2006} and Kerr nonlinearity
\cite{Schmidt1996,Hai2002,Niu2006,Sheng2011,Hamedi2015}. Using the slow light
greatly enhances the light-matter interaction and enables nonlinear optical
processes to achieve significant efficiency even at a single- photon level
\cite{Schmidt1996,Harris1998,Lukin2000,Wang2006,Shiau2011,Chen2012,Venkataraman2013,Maxwell2013,Baur2014}. 

There has been a considerable amount of activities on single-
\cite{Boller1991,Harris-PT-1997,Juzeliunas-02,Ruseckas2007,Ruseckas2011,Paspalakis2002,Fleischhauer2000,Fleischhauer2002,Lukin2003,Fleischhauer2005,Schnorrberger-PRL-2009}
and two-component (spinor)
\cite{Unanyan2010,Ruseckas2013,Lee2014,Fleischhauer2016,Ruseckas2011PRA,Bao2011coherent}
slow light in atomic media induced by the EIT. The former single-complonent
slow light involves a probe beam of light and one or several control beams
resonantly interacting with atomic media characterized by three level Lambda
($\Lambda$) type \cite{Harris-PT-1997,Ruseckas2007} or four level tripod type
\cite{Paspalakis2002,Raczynski-OC-2006,Raczynski-PRA-2007,Ruseckas2011,Mazets2017}
atom-light coupling schemes. In the later (spinor) case, double-tripod (DT)
\cite{Unanyan2010,Ruseckas2013,Lee2014,Fleischhauer2016,Ruseckas2011PRA,Bao2011coherent}
coupling schemes have been considered to support a simultaneous propagation of
two probe beams leading to formation of a two-component slow light. 

In this paper, we propose and analyze a novel five-level closed-loop scheme
supporting the EIT. Closed-loop quantum configurations
\cite{Payne2002,Hamedi2015,Ruseckas2013,Korsunsky1999,Shpaisman2005,Fleischhaker2008}
represent a class of atom-light coupling schemes in which the driving fields
acting on atoms build closed paths for the transitions between atomic levels.
The interference between different paths makes the system sensitive to relative
phases of the applied fields. In our proposal illustrated in figure~1, the
atom-light coupling represents a five-level combined Lambda-tripod scheme, in
which three atomic ground states are coupled to two excited states by four
control and one probe laser fields. In other words, the scheme involves four
atomic levels coupled between each other by four control fields and interacting
with a ground level through a weak probe field. The existence of dark states,
essential for the EIT, is analytically demonstrated for such an atom-light
coupling setup. It is shown that in some specific limiting cases this scheme
can be equivalent to the conventional $\Lambda$- or $N$-type atom-light
couplings. An advantage of such an atomic system is a possibility of
transitions to the $\Lambda$ or $N$-type level schemes just by changing the
amplitudes and phases of the control lasers. The limiting cases are discussed
where the scheme reduces to atom-light couplings of the $\Lambda$- or $N$-type.
By making a transition between the two limiting cases, one can switch from the
EIT regime to the absorption for the probe field propagation.

Laser-driven atomic media, on the other hand, can be exploited to exhibit
various nonlinear optical properties
\cite{Wu2003,Zhang2008,wu2004,Li2010,Joshi2003,Schmidt1996,Harris1998,Rajitha2015,Braje2004,Niu2006,Dey2007}.
A particular example is formation of optical solitons with applications for
optical buffers, phase shifters\cite{Kang2003}, switches\cite{Rodrigo2003},
routers, transmission lines \cite{Heebner2002}, wavelength converters
\cite{Melloni2003}, optical gates \cite{Liu2004} and others. Solitons represent
a specific type of stable shape-preserving waves propagating through nonlinear
media. They can be formed due to a balance between dispersive and nonlinear
effects leading to an undistorted propagation over long distance
\cite{Agrawal2001,Hasegawa2003,Liu2004,Xie2006,WuPRL2004,Burger1999,Denschlag2000,Huang2002,Kivshar1998,YuanYao2007,Xie2007}.
Following a report of ultraslow optical solitons in a highly resonant atomic
medium by Wu and Deng \cite{WuPRL2004}, these solitary waves have received a
considerable attention
\cite{YangPRA2009,WuOptLett2004,Hang2006,Si2009,Yang2011,ChaoHang2010,Chengjie2011,Si2010,Liu2010,Liang2010,Huang2005,Chen2015,Chen2014}.
Here, the equations of motion that govern the nonlinear evolution of the
probe-pulse envelope are derived for the Lambda-tripod atom-light coupling by
solving the coupled Maxwell-Bloch equations. It is found that, by properly
choosing the parameters of the system, the formation and slow propagation of
shape-preserving optical solitons is feasible.

The paper is organized as follows. In
section~\ref{sec:Formulation-and-theoretical} we present the proposed setup and
investigate propagation of the probe beam for various configurations of the
control fields. Nonlinear propagation of the probe pulse is considered in
section~\ref{sec:nonlinear}. Section~\ref{sec:concl} summarizes our findings.

\section{Formulation and theoretical background\label{sec:Formulation-and-theoretical}}

\subsection{The system}

\begin{figure}
\includegraphics[width=0.6\textwidth]{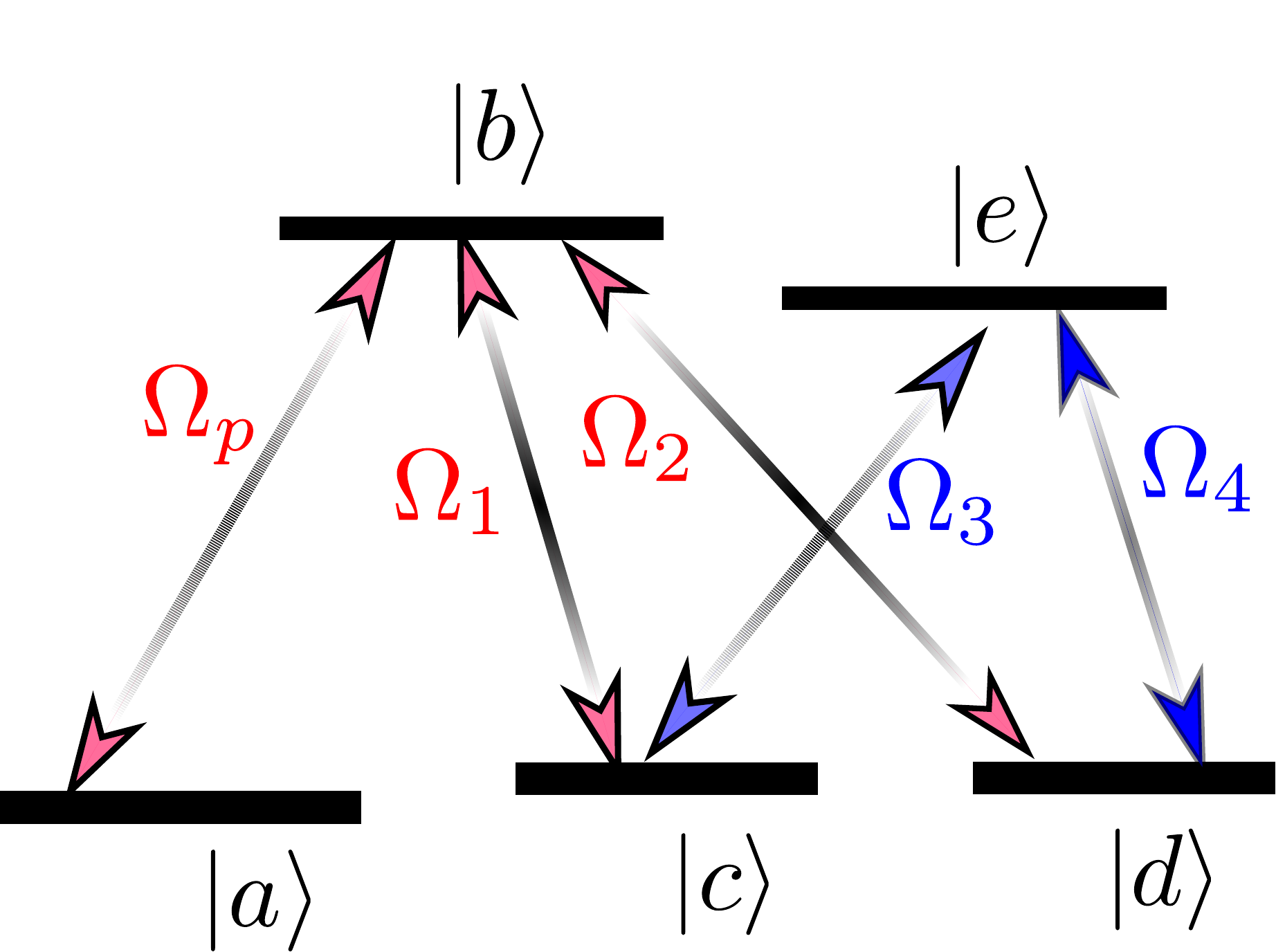}
\caption{Schematic diagram of the five-level Lambda-tripod quantum system.}
\label{fig:scheme}
\end{figure}

Let us consider a probe pulse described by a Rabi frequency $\Omega_{p}$.
Additional laser fields described by Rabi frequencies $\Omega_{1},$
$\Omega_{2}$, $\Omega_{3}$ and $\Omega_{4}$ control propagation of the probe
pulse. The probe and the control fields are assumed to co-propagate along the
$z$ direction. We shall analyze the light-matter interaction in an ensemble of
atoms using a five-level Lambda-tripod scheme shown in figure~\ref{fig:scheme}.
The atoms are characterized by three ground levels $|a\rangle$, $|c\rangle$ and
$|d\rangle$, as well as two excited states $|b\rangle$ and $|e\rangle$. Four
coherent control fields with the Rabi frequencies $\Omega_{1},$ $\Omega_{2}$,
$\Omega_{3}$ and $\Omega_{4}$ induce dipole-allowed transitions
$|b\rangle\longleftrightarrow|c\rangle$,
$|b\rangle\longleftrightarrow|d\rangle$,
$|e\rangle\longleftrightarrow|c\rangle$, and
$|e\rangle\longleftrightarrow|d\rangle$, respectively. As a result, the control
fields couple two excited states $|b\rangle$ and $|e\rangle$ via two different
pathways
$|b\rangle\stackrel{\Omega_{1}^{*}}{\rightarrow}|c\rangle\stackrel{\Omega_{3}}{\rightarrow}|e\rangle$
and
$|b\rangle\stackrel{\Omega_{2}^{*}}{\rightarrow}|d\rangle\stackrel{\Omega_{4}}{\rightarrow}|e\rangle$
making a four level closed-loop coherent coupling scheme described by the
Hamiltonian ($\hbar=1$) 
\begin{equation}
H_{\mathrm{4Levels}}=-\Omega_{1}^{*}|c\rangle\langle b|-\Omega_{2}^{*}|d\rangle\langle b|-\Omega_{3}^{*}|c\rangle\langle e|-\Omega_{4}^{*}|d\rangle\langle e|+\mathrm{H.c.}\,.
\label{eq:3}
\end{equation}
Furthermore, the tunable probe field with the Rabi frequency $\Omega_{p}$
induces a dipole-allowed optical transition $|a\rangle\longleftrightarrow|b\rangle$.
The total Hamiltonian of the system involving all five atomic levels
of the combined $\Lambda$ and tripod level scheme is given by 
\begin{equation}
H_{\mathrm{5Levels}}=-\left(\Omega_{p}^{*}|a\rangle\langle b|+\Omega_{p}|b\rangle\langle a|\right)+H_{\mathrm{4Levels}}\,.
\label{eq:4-1}
\end{equation}
Note that the complex Rabi frequencies of the four control fields can be
written as $\Omega_{j}=|\Omega_{j}|e^{i\phi_{i}}$, with $j=1,2,3,4$, where
$|\Omega_{j}|$ and $\phi_{j}$ are the amplitude and phase of each applied
field. As it will be explored below, in this scheme the destructive
interference between different transition pathways induced by the the control
and probe beams can make the medium transparent for the resonant probe beams in
a narrow frequency range due to the EIT. We define
$\phi=(\phi_{1}-\phi_{2})-(\phi_{3}-\phi_{4})$ to be a relative phase among the
four control fields forming a closed-loop coherent coupling. By changing
$\phi$, one can substantially modify the transparency and absorption properties
for the probe field in such a Lambda-tripod scheme. 

\subsection{Equations of motion}

The dynamics of the probe field propagating through the atomic medium is
described by the Maxwell-Bloch equations. To the first-order of $\Omega_{p}$
the equations have the form
\begin{eqnarray}
\dot{\rho}_{ba}^{(1)} =d_{1}\rho_{ba}^{(1)}+\rmi\Omega_{1}\rho_{ca}^{(1)}+
\rmi\Omega_{2}\rho_{da}^{(1)}+\rmi\Omega_{p},\label{eq:4}\\
\dot{\rho}_{ca}^{(1)} =d_{2}\rho_{ca}^{(1)}+\rmi\Omega_{1}^{*}\rho_{ba}^{(1)}+
\rmi\Omega_{3}^{*}\rho_{ea}^{(1)},\label{eq:5}\\
\dot{\rho}_{da}^{(1)} =d_{2}\rho_{da}^{(1)}+\rmi\Omega_{2}^{*}\rho_{ba}^{(1)}+
\rmi\Omega_{4}^{*}\rho_{ea}^{(1)},\label{eq:6}\\
\dot{\rho}_{ea}^{(1)} =d_{3}\rho_{ea}^{(1)}+\rmi\Omega_{3}\rho_{ca}^{(1)}+
\rmi\Omega_{4}\rho_{da}^{(1)},\label{eq:7}
\end{eqnarray}
and 
\begin{equation}
\frac{\partial\Omega_{p}}{\partial z}+c^{-1}\frac{\partial\Omega_{p}}{\partial t}=
\rmi\eta\rho_{ba}^{(1)}\,,\quad\mathrm{with}\quad\eta=
\frac{2N\omega_{p}|\mu_{ba}|^{2}}{\hbar c}\,,\label{eq:8}
\end{equation}
where $\rho_{u,v}^{(1)}$ are the first-order matrix elements of the density
materix operator $\rho=\sum|u\rangle\rho_{uv}\langle v|$.  The optical Bloch
equations (\ref{eq:4})--(\ref{eq:7}) imply the probe field to be much weaker
than the control ones. In that case most atomic population is in the ground
state $|a\rangle$, and one can treat the probe field as a perturbation.
Therefore, we can apply the perturbation expansion
$\rho_{ij}=\Sigma_{k}\rho_{ij}^{(k)},$ where $\rho_{ij}^{(k)}$ represents the
$k$th order part of $\rho_{ij}$ in terms of probe field $\Omega_{p}$. Since
$\Omega_{p}\ll\Omega_{i}$ $(i=1,2,3,4)$, the zeroth-order solution is
$\rho_{aa}^{(0)}=1$, while other elements being zero
($\rho_{bb}^{(0)}=\rho_{cc}^{(0)}=\rho_{dd}^{(0)}=\rho_{ee}^{(0)}=0$).  All
fast-oscillating exponential factors associated with central frequencies and
wave vectors have been eliminated from the equations, and only the
slowly-varying amplitudes are retained. 

The wave equation (\ref{eq:8}) describes propagation of the probe field
$\Omega_{p}$ influenced by the atomic medium, where $\mu_{ba}$ is an electric
dipole matrix element corresponding to the transition
$|b\rangle\longleftrightarrow|a\rangle$, $N$ is the atomic density and
$\omega_{p}$ is the frequency of the probe field. The density matrix equations
(\ref{eq:4})\textendash (\ref{eq:7}) describe the evolution of the atomic
system affected by the control and probe fields.  They follow from the general
quantum Liouville equation for the density matrix operator
\cite{Scully-Book-1997}
\begin{equation}
\dot{\rho}=-\frac{\rmi}{\hbar}[H_{\mathrm{5Levels}},\rho]+L_{\rho},\label{eq:9}
\end{equation}
where the damping operator $L_{\rho}$ describes the decay of the system
described by parameters $d_{1}=-\Gamma_{b}/2+\rmi\Delta_{p}$,
$d_{2}=\rmi(\Delta_{p}-\Delta_{2})$ and
$d_{3}=-\Gamma_{e}/2+\rmi(\Delta_{p}+\Delta_{3}-\Delta_{2})$ in
equations~(\ref{eq:4})--(\ref{eq:7}). We have defined the detunings as:
$\Delta_{2}=\Delta_{bc}=\Delta_{bd},$ and $\Delta_{3}=\Delta_{ec}=\Delta_{ed}$,
with $\Delta_{bc}=\omega_{1}-\omega_{bc}$,
$\Delta_{bd}=\omega_{2}-\omega_{bd}$, $\Delta_{ec}=\omega_{3}-\omega_{ec}$,
$\Delta_{ed}=\omega_{4}-\omega_{ed}$, and $\Delta_{p}=\omega_{p}-\omega_{ba}$,
where $\omega_{i}$ is a central frequency of the corresponding control field.
Two excited states $\left|b\right\rangle $ and $\left|e\right\rangle $ decay
with rates $\Gamma_{b}$ and $\Gamma_{e}$, respectively. 

\subsection{Transition to a new basis}

The Hamiltonian for the atomic four-level subsystem (\ref{eq:3})
can be represented as 
\begin{equation}
H_{\mathrm{4Levels}}=-\beta|D_{e}\rangle\langle b|-\alpha|B_{e}\rangle\langle b|-
\Omega|B_{e}\rangle\langle e|+\mathrm{H.c.},\label{eq:10}
\end{equation}
where
\begin{eqnarray}
|D_{e}\rangle =\frac{1}{\Omega}\left(\Omega_{4}|c\rangle-
  \Omega_{3}|d\rangle\right),\label{eq:D_e}\\
|B_{e}\rangle =\frac{1}{\Omega}\left(\Omega_{3}^{*}|c\rangle+
  \Omega_{4}^{*}|d\rangle\right)\label{eq:B_e}
\end{eqnarray}
are the internal dark and bright states for the $\Lambda$- scheme
made of the two ground states states $|c\rangle$ and $|d\rangle$,
as well as an excited states $|e\rangle$. One can also introduce
another set of dark and bright states corresponding to the $\Lambda$-
scheme made of the same pair of ground states states $|c\rangle$
and $|d\rangle$, yet a different excited state $|b\rangle$: 
\begin{eqnarray}
|D_{b}\rangle =\frac{1}{\Omega}\left(\Omega_{2}|c\rangle
  -\Omega_{1}|d\rangle\right),\label{eq:D_b}\\
|B_{b}\rangle =\frac{1}{\Omega}\left(\Omega_{1}^{*}|c\rangle
  +\Omega_{2}^{*}|d\rangle\right).\label{eq:B_b}
\end{eqnarray}
In writing equation~(\ref{eq:10}), the coefficient
\begin{equation}
\beta=\langle D_{b}|B_{e}\rangle=\langle D_{e}|B_{d}\rangle=
\frac{1}{\Omega}(\Omega_{1}^{*}\Omega_{4}^{*}-\Omega_{2}^{*}\Omega_{3}^{*})
\label{eq:beta}
\end{equation}
represents the quantum inteference between the four control fields
playing the main role in tuning dispersion and absorption properties
in the combined tripod and $\Lambda$ scheme. In addition, we define
\begin{equation}
\alpha=\langle D_{b}|D_{e}\rangle=\langle B_{e}|B_{d}\rangle=
\frac{1}{\Omega}(\Omega_{1}^{*}\Omega_{3}+\Omega_{2}^{*}\Omega_{4})
\label{eq:alpha}
\end{equation}
and the total Rabi frequency 
\begin{equation}
\Omega=\sqrt{|\Omega_{3}|^{2}+|\Omega_{4}|^{2}}\,.\label{eq:13}
\end{equation}
By changing the quantum interference coefficient $\beta$ and the
coefficient $\alpha$ one arrives at three different situations. For
each of them we shall plot the level schemes in the basis involving
the transformed states $|B_{e}\rangle$ and $|D_{e}\rangle$.

\subsubsection{Situation (a): $\beta\protect\neq0$ and $\alpha\protect\neq0$}

\begin{figure}
\includegraphics[width=0.6\textwidth]{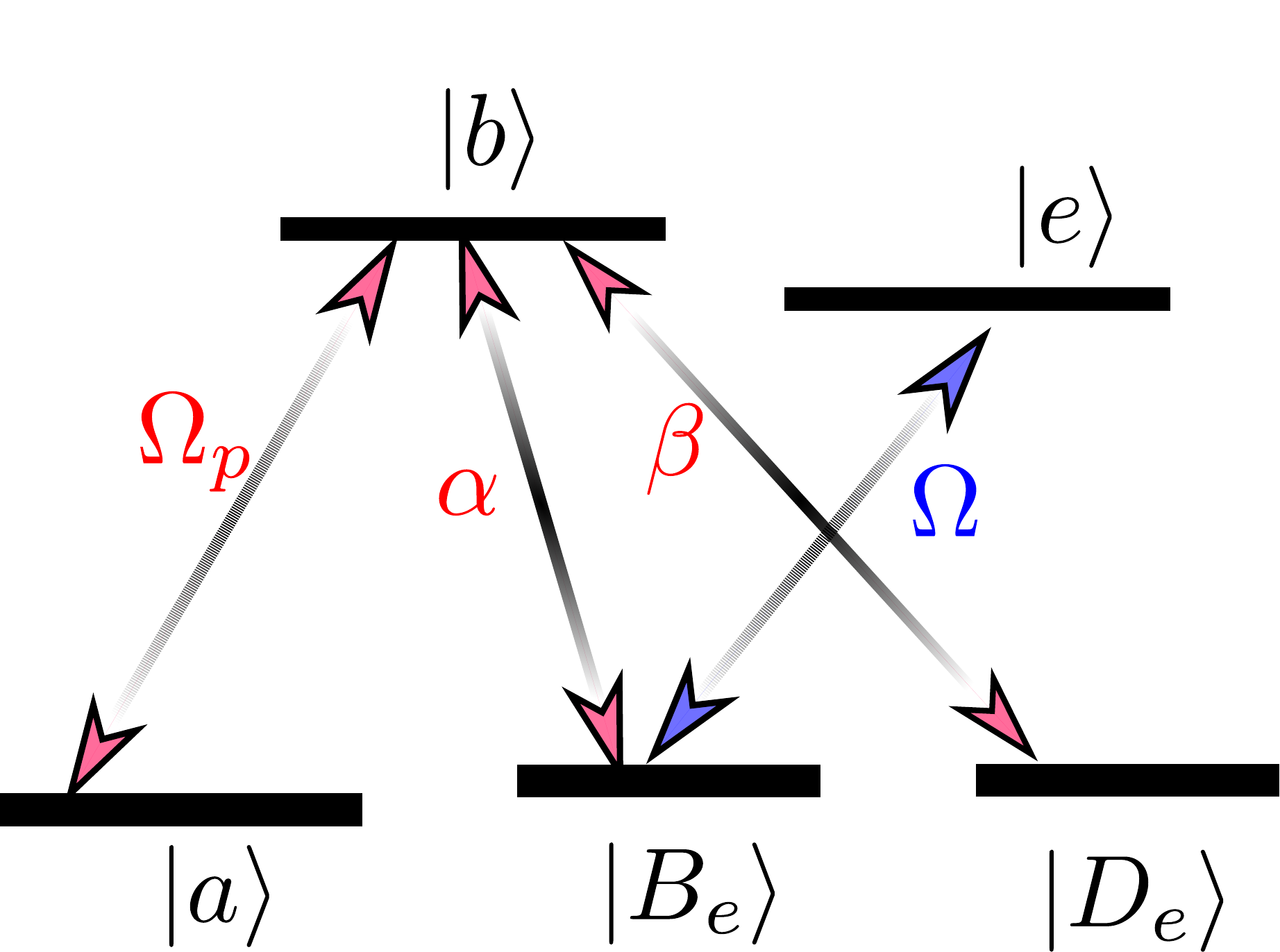}
\caption{Five-level quantum system in the transformed basis for $\beta\protect\neq0$
and $\alpha\protect\neq0$.}
\label{fig:scheme-a}
\end{figure}

In the case when both coefficients $\alpha$ and $\beta$ are nonzero, the
five-level tripod and $\Lambda$ scheme shown in figure~\ref{fig:scheme-a} looks
similar to the original scheme (figure~\ref{fig:scheme}) in the transformed
basis, but the coupling between the states $|D_{e}\rangle$ and $|e\rangle$ is
missing. The coefficient $\beta$ is nonzero when
$\Omega_{1}\Omega_{4}\neq\Omega_{2}\Omega_{3}$. The condition $\beta\neq0$ is
valid provided $|\Omega_{1}||\Omega_{4}|\neq|\Omega_{2}||\Omega_{3}|$ and phase
$\phi$ is arbitrary, or $|\Omega_{1}||\Omega_{4}|=|\Omega_{2}||\Omega_{3}|$
with $\phi\neq0$. 

When both $\alpha$ and $\beta$ are nonzero, one can define a global dark state
$|D\rangle$ for the whole atom-light coupling scheme 
\begin{equation}
|D\rangle=\beta|a\rangle-\Omega_{p}|D_{e}\rangle\,.\label{eq:16}
\end{equation}
The dark state is an eigenstate of the full atom-light Hamiltonian with a zero
eigen-energy: $H_{\mathrm{5Levels}}|D\rangle=0$. The state $|D\rangle$ has no
contribution by the ground state superpositon $|B_{e}\rangle$, as well as no
contribution by the bare excited states $|e\rangle$ and $|b\rangle$. As a
result, there is no transition from the state $|D\rangle$ to the excited states
$|e\rangle$ and $|b\rangle$, making the five-level closed- loop scheme
transparent to the electromagnetic field. This is a new mechanism for the EIT
compared with the $\Lambda$ \cite{Harris-PT-1997,Fleischhauer2000}, tripod
\cite{Paspalakis2002,Ruseckas2011}, or double tripod schemes
\cite{Ruseckas2013,Lee2014}.

It is known that the real and imaginary parts of $\rho_{ba}^{(1)}$ correspond
to the probe dispersion and absorption, respectively. A steady-state solution
to the density matrix element $\rho_{ba}$ reads under the resonance condition
$\Delta_{2}=\Delta_{3}=0$ 
\begin{eqnarray}
  \fl\rho_{ba}^{(1)}=\Omega_{p}\Delta_{p}(|\Omega_{3}|^{2}+
    |\Omega_{4}|^{2}+\rmi\Delta_{p}(-\Gamma_{e}/2
      +\rmi\Delta_{p}))\nonumber\\
      \times\left[|\beta|^{2}\Omega^{2}
  +\rmi\Delta_{p}(-\Gamma_{e}/2
    +\rmi\Delta_{p})(|\Omega_{1}|^{2}+|\Omega_{2}|^{2})\right.\nonumber\\
  +\left.\rmi\Delta_{p}(-\Gamma_{b}/2
    +\rmi\Delta_{p})\left(|\Omega_{3}|^{2}+|\Omega_{4}|^{2}
    +\rmi\Delta_{p}(-\Gamma_{e}/2+\rmi\Delta_{p})\right)\right]^{-1},
\label{eq:17}
\end{eqnarray}
where the interference term $\beta$ is involved.

A denominator of equation~(\ref{eq:17}) represents the fourth order polynomial
which contains zero points at four different detunings $\Delta_{p}$ of the
probe field from the EIT resonance. This provides four maxima in the absorption
profile of the system, as one can see in figure~\ref{fig:probe-absorption}(a).
Furthermore the EIT window is formed for zero detuning. In
\ref{sec:appendix-A} we have presented eigenstates and the
corresponding eigenvalues of the Hamiltonian~(\ref{eq:10}) describing the
four-level subsystem.  One can see that all four eigenstates $|n_{i}\rangle$
characterized by the eigenvalues $\lambda_{1,2}=\mp\frac{\sqrt{S-Y}}{\sqrt{2}}$
and $\lambda_{3,4}=\mp\frac{\sqrt{S+Y}}{\sqrt{2}}$ contain contributions due to
the excited state $|b\rangle$ (note that $S$ and $Y$ can be found in
\ref{sec:appendix-A}). This results in four peaks in absorption
profile of the system. 

\subsubsection{Situation (b): $\beta=0$ and $\alpha\protect\neq0$}

\begin{figure}
\includegraphics[width=0.6\textwidth]{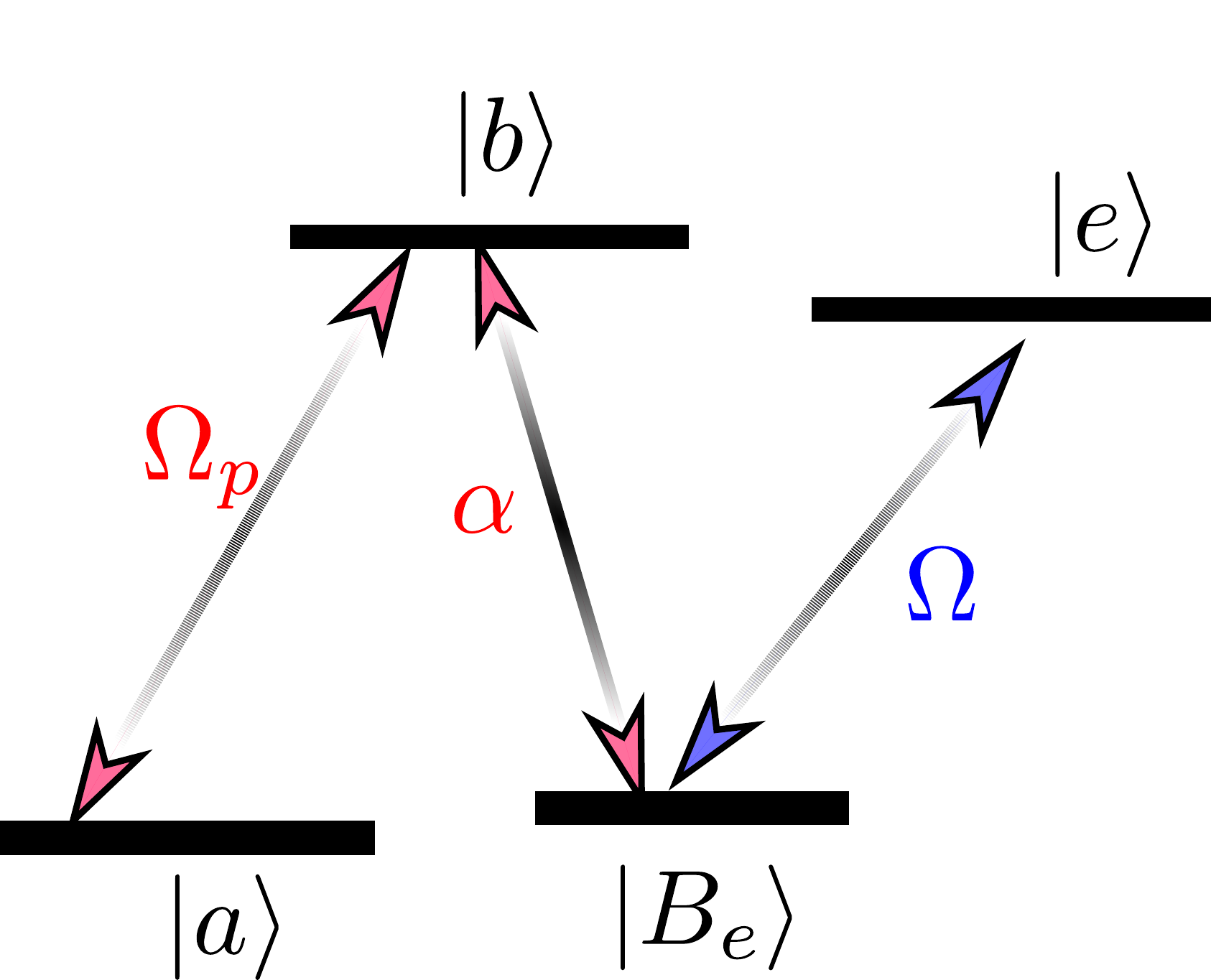} 
\caption{The level scheme in the transformed basis for $\beta=0$ and
  $\alpha\protect\neq0$.  The ground state superposition $|D_{e}\rangle$ (now
  shown in the figure) is decoupled from the remaining four states.}
\label{fig:scheme-b}
\end{figure}

The condition $\beta=0$ is fulfilled if
$\Omega_{1}\Omega_{4}=\Omega_{2}\Omega_{3}$, or equivalently
$|\Omega_{2}||\Omega_{3}|=|\Omega_{1}||\Omega_{4}|$ and $\phi=0$. In that case
the state $|D_{e}\rangle$ is not involved, so the interaction
Hamiltonian~(\ref{eq:10}) for the four-level subsystem can be rewritten as
\begin{equation}
H_{\mathrm{4Levels}}=-\alpha|B_{e}\rangle\langle b|-
\Omega|B_{e}\rangle\langle e|+\mathrm{H.c.}\,.
\label{eq:18}
\end{equation}
Consequently the five-level tripod and $\Lambda$ scheme becomes equivalent
to a conventional $N$-type atomic system \cite{Sheng2011,Harris1999}
shown in figure~\ref{fig:scheme-b}.

Since the quantum interference term $\beta$ vanishes, equation~(\ref{eq:17})
simplifies to 
\begin{eqnarray}
  \fl\rho_{ba}^{(1)}=\Omega_{p}(|\Omega_{3}|^{2}+|\Omega_{4}|^{2}
    +\rmi\Delta_{p}(-\Gamma_{e}/2
      +\rmi\Delta_{p}))\nonumber\\
      \times\left[\rmi (-\Gamma_{e}/2
    +\rmi\Delta_{p})(|\Omega_{1}|^{2}+|\Omega_{2}|^{2})\right.\nonumber\\
  +\left.\rmi (-\Gamma_{b}/2+\rmi\Delta_{p})(|\Omega_{3}|^{2}
    +|\Omega_{4}|^{2}+\rmi\Delta_{p}(-\Gamma_{e}/2
      +\rmi\Delta_{p}))\right]^{-1}.
\label{eq:19}
\end{eqnarray}
The denominator of $\rho_{ba}^{(1)}$ in equation~(\ref{eq:19}) is now a cubic
polynomial, providing three absorption maxima. The eigenstates and eigenvalues
corresponding to this situation are presented in \ref{sec:appendix-B}.
The eigenvector $|n_{2}\rangle$ characterized by a zero eigen-energy coincides
with the dark state $|D_{e}\rangle$ and is decoupled from the radiation fields.
Only the remaining three eigenvectors $|n_{1}\rangle$, $|n_{3}\rangle$ and
$|n_{4}\rangle$ contain the contribution due to an excited state $|b\rangle$.
This leads to three absorption peaks displayed in
figure~\ref{fig:probe-absorption}(b). In this way, the absence of quantum
interference term $\beta$ between the control fields destroys one of the peaks
in the absorption profile leading to three absorption maxima. 

\subsubsection{Situation (c): $\beta\protect\neq0$ and $\alpha=0$}

\begin{figure}
\includegraphics[width=0.6\textwidth]{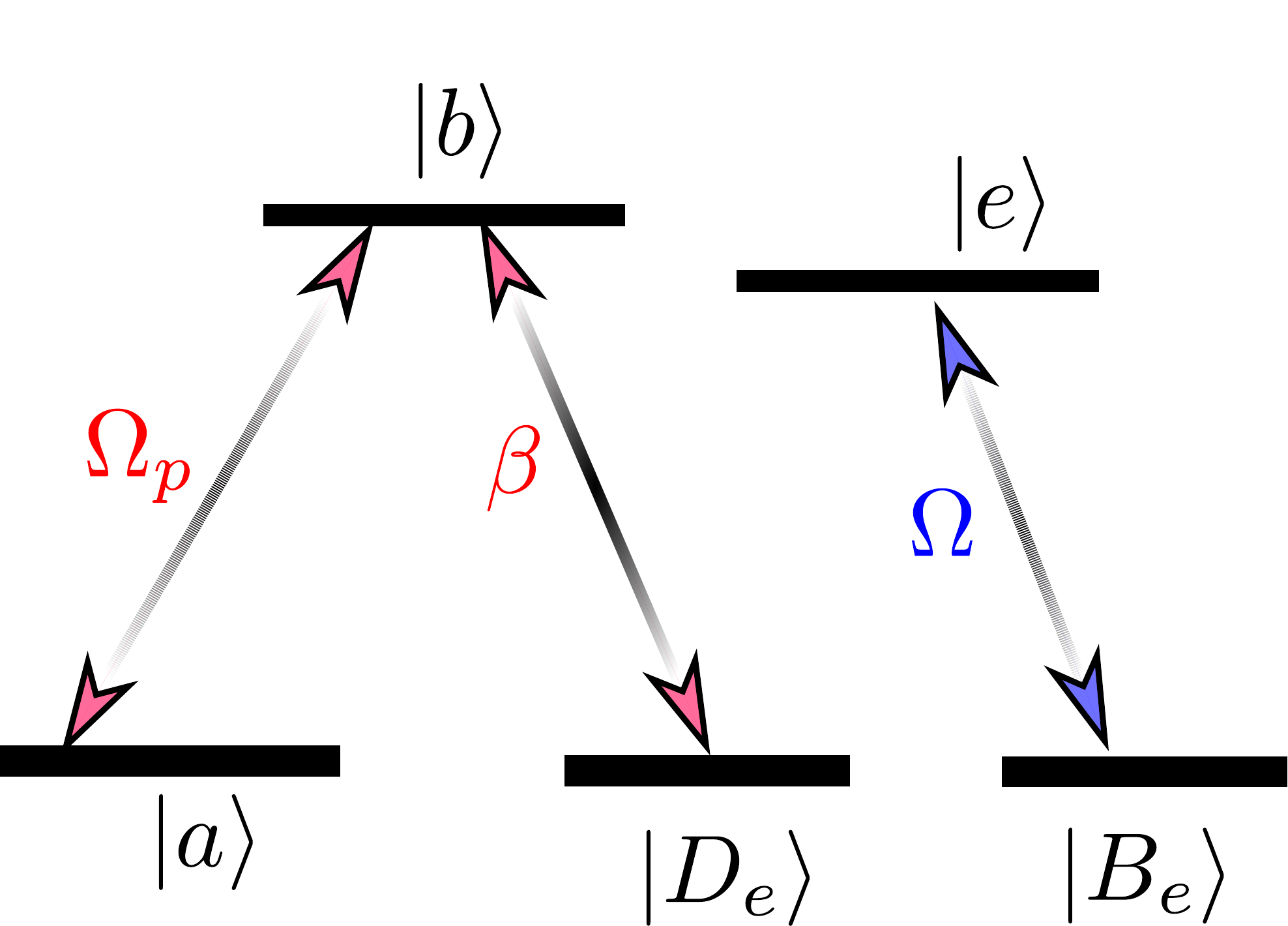} 
\caption{Schematic diagram of the five-level quantum system for
  $\beta\protect\neq0$ and $\alpha=0$.}
\label{fig:scheme-c}
\end{figure}

When the coefficient $\beta$ is nonzero but the coefficient $\alpha$
is zero, the interaction Hamiltonian~(\ref{eq:10}) can be represented
as
\begin{equation}
H_{\mathrm{4Levels}}=-\beta|D_{e}\rangle\langle b|
-\Omega|B_{e}\rangle\langle e|+\mathrm{H.c.}\,.\label{eq:20}
\end{equation}
As illustrated in figure~\ref{fig:scheme-c}, the five-level tripod and
$\Lambda$ scheme is then equivalent to a conventional $\Lambda$-type atomic
system \cite{Harris-PT-1997,Ruseckas2007} which is decoupled from the two-level
system involving the states $|B_{e}\rangle$ and $|e\rangle$. 

In the following we consider a symmetric case where
$|\Omega_{1}|=|\Omega_{2}|$, $|\Omega_{3}|=|\Omega_{4}|$ and $\phi=\pi$. In
such a situation the conditions $\beta\neq0$ and $\alpha=0$ are fulfilled, with
$\beta^{2}=2|\Omega_{1}^{2}|$.  Equation~(\ref{eq:17}) for $\rho_{ba}$ then
simplifies considerably, giving
\begin{equation}
\rho_{ba}^{(1)}=\Omega_{p}\frac{\Delta_{p}}{|\beta|^{2}
  +\rmi\Delta_{p}\left(-\frac{\Gamma_{b}}{2}+\rmi\Delta_{p}\right)}\,.
\label{eq:21}
\end{equation}
Obviously, the polynominal in denominator becomes quadratic in $\Delta_{p}$
resulting in two absorption peaks or a single EIT window, which is a
characteristic feature of the $\Lambda$ scheme. Furthermore, the probe
absorption and dispersion do not dependent on $\Omega$, only $\beta$
contributes to the optical properties. This is because the $\Lambda$-type
scheme is now decoupled from the transition
$|B_{e}\rangle\stackrel{\Omega}{\rightarrow}|e\rangle$, and the system behaves
as a three- level $\Lambda$-type scheme containing $|a\rangle,$ $|b\rangle$,
and $|D_{e}\rangle$. 

\begin{figure}
\includegraphics[width=0.33\textwidth]{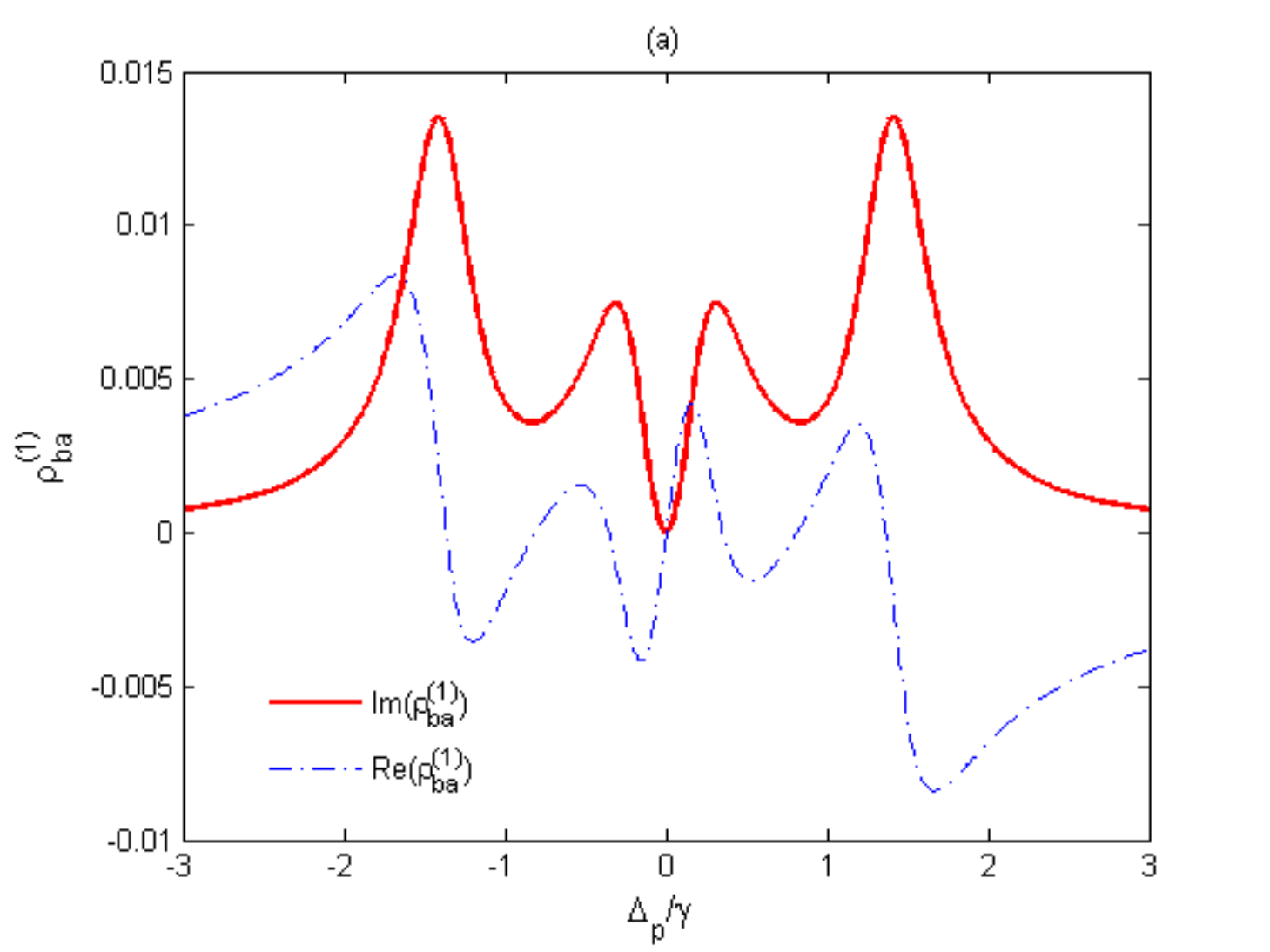} \includegraphics[width=0.33\textwidth]{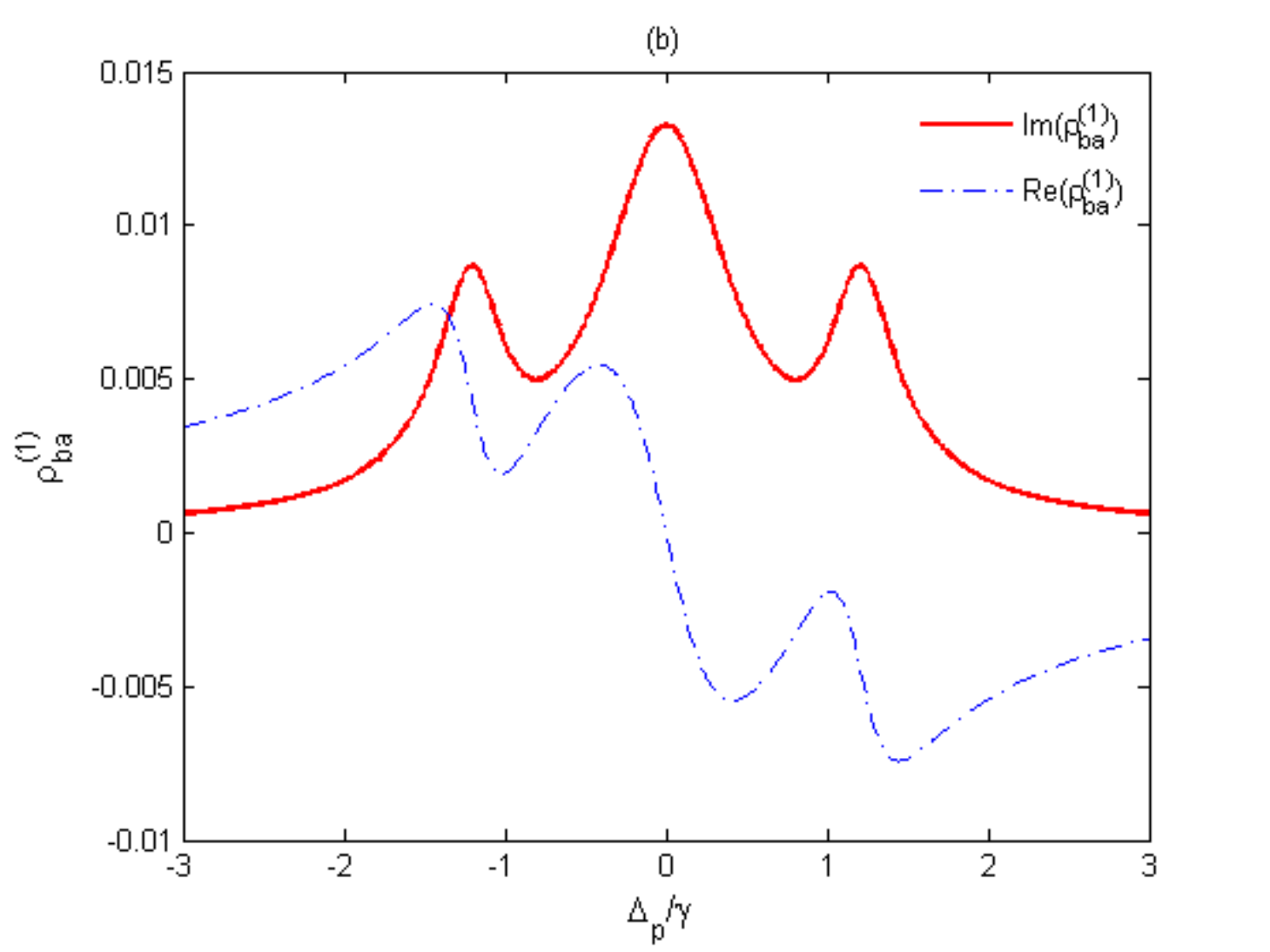}\includegraphics[width=0.33\textwidth]{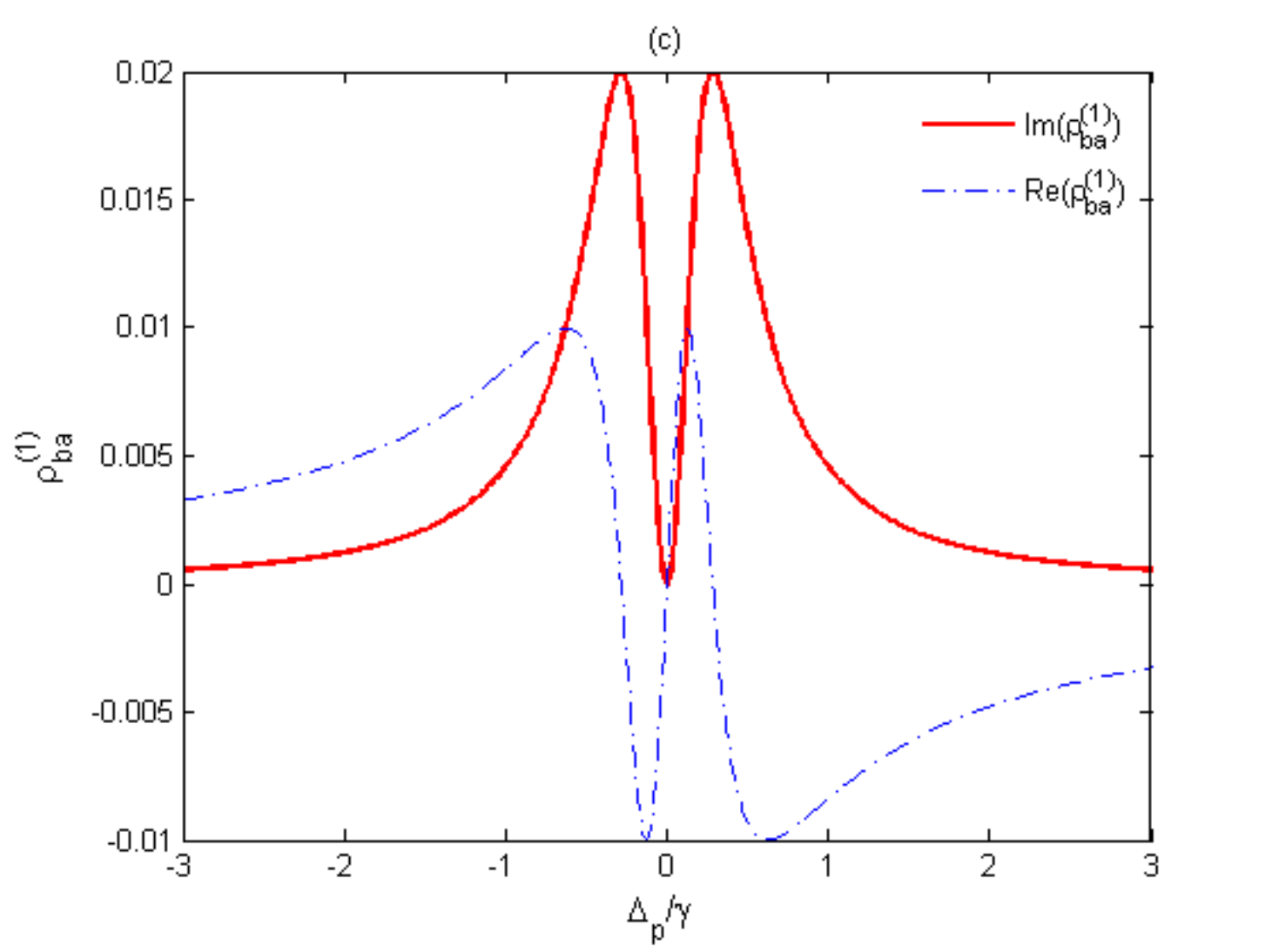}
\caption{Probe absorption ($\mathrm{Im}(\rho_{ba}^{(1)}$)) versus $\Delta_{p}$
  for (a) $\Omega_{1}=0.9\gamma$, $\Omega_{2}=0.7\gamma$,
  $\Omega_{3}=0.4\gamma$, $\Omega_{4}=0.8\gamma$, and $\phi=0$ corresponding to
  the first situation, (b) $\Omega_{1}=\Omega_{2}=0.5\gamma$,
  $\Omega_{3}=\Omega_{4}=0.7\gamma$ and $\phi=0$ corresponding to the second
  situation, and (c) $\Omega_{1}=\Omega_{2}=0.2\gamma$,
  $\Omega_{3}=\Omega_{4}=0.1\gamma$ and $\phi=\pi$ corresponding to the third
  situation. Other parameters are $\Gamma_{e}=\Gamma_{b}=\gamma$,
  $\Delta_{2}=\Delta_{3}=0$, and $\Omega_{p}=0.01\gamma$. Note that all
  frequencies are scaled by $\gamma$ which should be in the order of MHz, like
  for cesium (Cs) atoms.}
\label{fig:probe-absorption}
\end{figure}

In the following we summarize our results for the behavior of real and
imaginary parts of $\rho_{ba}^{(1)}$ corresponding to the probe dispersion and
absorption for different situations (a)\textendash (c) described above. Without
a loss of generality, in all the simulations we take
$\Gamma_{e}=\Gamma_{b}=\gamma$. The other frequencies are scaled by $\gamma$
which should be in the order of MHz for cesium (Cs) atoms.
Figure~\ref{fig:probe-absorption} shows that the our model provides a high
control of dispersive-absorptive optical properties of the probe field. The
absorption profile has four, three and two peaks featured in
figures~\ref{fig:probe-absorption}(a)--\ref{fig:probe-absorption}(c),
respectively. There is the resulting change in the sign of the slope of the
dispersion at $\Delta_{p}=0$ for different situations (a)--(c).  This
gives rise to switching in the group velocity of the probe pulse from
subluminal to superluminal or visa versa. In particular, for the choice of
parameters satisfying situation (a) and (c), there is the subluminality
accompanied by EIT at line center. On the other hand, the superluminality
accompanied by a considerable absorption is observed for the parametric
condition satisfying the situation (b).

\section{Linear and nonlinear pulse propagation in combined tripod and $\Lambda$
scheme\label{sec:nonlinear}}

In this section we consider propagation of the probe pulse in the proposed
tripod and $\Lambda$ scheme. Performing the time Fourier transform of
equations~(\ref{eq:4})--(\ref{eq:8}) one can obtain
\begin{eqnarray}
t_{1}(\omega)F_{ba}+\Omega_{1}F_{ca}+\Omega_{2}F_{da}+\Lambda_{p} =0,\label{eq:22}\\
t_{2}(\omega)F_{ca}+\Omega_{1}^{*}F_{ba}+\Omega_{3}^{*}F_{ea} =0,\label{eq:23}\\
t_{2}(\omega)F_{da}+\Omega_{2}^{*}F_{ba}+\Omega_{4}^{*}F_{ea} =0,\label{eq:24}\\
t_{3}(\omega)F_{ea}+\Omega_{3}F_{ca}+\Omega_{4}F_{da} =0,\label{eq:25}
\end{eqnarray}
and 
\begin{equation}
\frac{\partial\Lambda_{p}}{\partial z}-\rmi\frac{\omega}{c}\Lambda_{p}=
\rmi\eta F_{ba},\label{eq:26}
\end{equation}
where $t_{1}(\omega)=\omega+i\Gamma_{b}/2+\Delta_{p}$,
$t_{2}(\omega)=\omega+\Delta_{p}-\Delta_{2}$, and
$t_{3}(\omega)=\omega+\rmi\Gamma_{e}/2+\Delta_{p}-\Delta_{2}+\Delta_{3}$.  Note
that $F_{ij}$ and $\Lambda_{p}$ represent the Fourier transforms of
$\rho_{ij}^{(1)}$ and $\Omega_{p}$, respectively, where $\omega$ is a deviation
from the central frequency. 

A solution of equation~(\ref{eq:26}) is a plane wave of the form
\begin{equation}
\Lambda_{p}(z,\omega)=\Lambda_{p}(0,\omega)\rme^{\rmi\kappa(\omega)z},\label{eq:27}
\end{equation}
where
\begin{equation}
\kappa=\frac{\omega}{c}-\frac{\eta S_{1}(\omega)}{Q(\omega)},\label{eq:28}
\end{equation}
describes the linear dispersion relation of the system. Expanding $\kappa$ in
power series around the center frequency of the probe pulse ($\omega=0$) and
taking only the first three terms, we get
\begin{equation}
\kappa=\kappa_{0}+\kappa_{1}\omega+\kappa_{2}\omega^{2},\label{eq:29}
\end{equation}
where the detaied expressions for the coefficients $\kappa_{0}$, $\kappa_{1}$
and $\kappa_{2}$ are given in \ref{sec:appendix-C}, while
$S_{1}(\omega)$ and $Q(\omega)$ can be found in \ref{sec:appendix-D}.
In equation~(\ref{eq:29}),
$\kappa_{i}=\frac{d^{j}\kappa(\omega)}{d\omega^{j}}|_{\omega=0}$ with $j=0,1,2$
are the dispersion coefficients in different orders.  In general, the real part
of $\kappa_{0}=\Upsilon+\rmi\chi/2$ defines the phase shift $\Upsilon$ per
unit length, while the imaginary part indicates the linear absorption $\chi$ of
the probe pulse. The group velocity $v_{g}$ is given by $1/\kappa_{1}$, whereas
the quadratic term $\kappa_{2}$ is associated with the group velocity
dispersion which causes the pulse distortion. 

In the linear regime, we take an incoming probe pulse to be of the Gausian
shape, $\Omega_{p}(0,t)=\Omega_{p}^{0}\rme^{-(t/\tau_{0})^{2}}$, with a duration
$\tau_{0}$. The subsequent time evolution is obtained from
equation~(\ref{eq:27}) by carrying out an inverse Fourier transform
\cite{WuPRL2004} 
\begin{equation}
\Omega_{p}(z,t)=\frac{\Omega_{p}^{0}}{\sqrt{L_{z}}}
\exp\left[\rmi\kappa_{0}z-\frac{(t-\kappa_{1}z)^{2}}{L_{z}\tau_{0}^{2}}\right]\,,
\label{eq:30}
\end{equation}
with $L_{z}=s_{1}(z)-\rmi s_{2}(z)$,
$s_{1}(z)=1+4z\mathrm{Re}(\kappa_{2})/\tau_{0}^{2}$ and
$s_{2}(z)=4z\mathrm{Im}(\kappa_{2})/\tau_{0}^{2}$ . In this way, even if there
is no absorption due to EIT ($\mathrm{Im}(\kappa_{i})=0$, $i=0,1,2$), the
dispersion effects can contribute to the pulse attenuation and spreading during
propagation.

Our goal is to obtain shape-preserving optical pulses which can propagate
without significant distortion and loss in our medium. The idea is to include
the optical Kerr nonlinearity of the probe laser field into the light
propagation, and show that the Kerr nonlinear effect can compensate the
dispersion effects and result in shape-preserving optical pulses. To balance
the dispersion effects and optical nonlinearity, in the following a theoretical
model is employed based on the coupled Maxwell-Bloch equations for the
nonlinear pulse propagation. Following \cite{WuPRL2004}, we take a trial
function
\begin{equation}
\Lambda_{p}=\tilde{\Lambda}_{p}\rme^{\rmi\kappa_{0}z}\,.\label{eq:32}
\end{equation}
Substituting equation~(\ref{eq:32}) into the wave equation (\ref{eq:26})
and using equations~(\ref{eq:28}), (\ref{eq:29}) and (\ref{eq:77})
we obtain
\begin{equation}
\frac{\partial\Lambda_{p}}{\partial z}\rme^{\rmi\kappa_{0}z}=
\rmi(\kappa_{1}\omega+\kappa_{2}\omega^{2})\Lambda_{p}\rme^{\rmi\kappa_{0}z},
\label{eq:33}
\end{equation}
where we have replaced $\tilde{\Lambda}_{p}$ with $\Lambda_{p}$ for the sake of
convenience. Here we only keep terms up to the order $\omega^{2}$ in expanding
the dispersion relarion $\kappa_{j}$. 

In deriving the linearized wave equation (\ref{eq:8}), the nonlinear
polarization due to the optical Kerr nonlinearity of the probe field has been
neglected. Now we turn to investigate the nonlinear propagation of light due to
the Kerr effect. To incorporate the nonlinear optical terms in the pulse
propagation, the right hand side of wave equation (\ref{eq:8}) must be
rewritten as $\rmi\eta\rho_{ba}^{(1)}-\rmi\eta\mathrm{Kerr}$.  The Kerr nonlinear
term has an opposite sign than the linear term $\rmi\eta\rho_{ba}^{(1)}$ and the
probe absorption and dispersion are proportional to imaginary and real parts of
$\rho_{ba}^{(1)}$, respectively.  Consequently a large optical nonlinearity can
cancel the dispersion and suppress the absorption of probe field, effectively.
A derivation of the Kerr nonlinear coefficient is provided in
\ref{sec:appendix-E}.

Performing the inverse Fourier transform of equation~(\ref{eq:33}), using the
expression (\ref{eq:36b-1}) for $\mathrm{Kerr}$ and introducing new coordinates
$\zeta=z,$ and $\eta=t-z/v_{g},$ we arrive at the nonlinear wave equation for
the slowly varing envelope $\Omega_{p}$ 
\begin{equation}
  \rmi\frac{\partial}{\partial\zeta}\Omega_{p}-
  \kappa_{2}\frac{\partial^{2}}{\partial\eta^{2}}\Omega_{p}=
  \Theta \rme^{-\chi\zeta}|\Omega_{p}|^{2}\Omega_{p}.
  \label{eq:37-2}
\end{equation}
where $\Theta=-\eta\mathrm{Kerr}$, and
$\chi=2\mathrm{Im}(\kappa_{0})=2\eta\mathrm{Im}\left(\frac{-S_{1}}{Q}\right)$.

Equation (\ref{eq:37-2}) contains generally complex coefficients.  However, for
suitable set of system parameters, the absorption coefficent $\chi$ may be very
small, i.e., $\chi\simeq0$, and imaginary parts of coefficients $\Theta$ and
$K_{2}$ may be made very small in comparison to their real parts, i.e.,
$\kappa_{2}=\kappa_{2r}+\rmi\kappa_{2i}\approx\kappa_{2r}$, and
$\Theta=\Theta_{r}+\rmi\Theta_{i}\approx\Theta_{r}$. In this case
equation~(\ref{eq:37-2}) can be written as
\begin{equation}
  \rmi\frac{\partial}{\partial\zeta}\Omega_{p}-
  \kappa_{2r}\frac{\partial^{2}}{\partial\eta^{2}}\Omega_{p}=
  \Theta_{r}|\Omega_{p}|^{2}\Omega_{p}.
  \label{eq:40-2}
\end{equation}
This represents the conventional nonlinear Schrodinger (NLS) equation which
describes the nonlinear evolution of the probe pulse and allows bright and dark
soliton solutions. The nature of the soliton solution is determined by a sign of
the product $\kappa_{2r}\Theta_{r}$. A bright soliton is obtained for
$\kappa_{2r}\Theta_{r}>0$, and the solution is then given by
\begin{equation}
\Omega_{p}=\Omega_{p0}\mathrm{sech}(\eta/\tau)
\exp\left(-\rmi\zeta\Theta_{r}|\Omega_{p}|^{2}/2\right)\,.
\end{equation}
For $\kappa_{2r}\Theta_{r}<0$, one obtains the dark solion solution of the form
\begin{equation}
\Omega_{p}=\Omega_{p0}\mathrm{tanh}(\eta/\tau)
\exp\left(-\rmi\zeta\Theta_{r}|\Omega_{p}|^{2}/2\right)\,.
\end{equation}
Here $\Omega_{p0}=(1/\tau)\sqrt{|\kappa_{2r}/\Theta_{r}|}$ represents an
amplitude of the probe field and $\tau$ is the typical pulse duration (soliton
width).

In the following, we explore a possibility for the formation of the shape
preserving optical solitons in this combined tripod and $\Lambda$ scheme for a
realistic atomic system and present numeric calculations.  The proposed scheme
involving the five-level combined tripod and $\Lambda$ structure can be
experimentally implemented using the cesium (Cs) atom vapor. In our proposal,
the levels $|c\rangle$, $|d\rangle$ and $|e\rangle$ can correspond to
$|6S_{1/2},F=3,M_{F}=+1\rangle$, $|6S_{1/2},F=3,M_{F}=+3\rangle$ and
$|6P_{3/2},F=2,M_{F}=+2\rangle$, respectively. In addition, the levels
$|b\rangle$ and $|a\rangle$ can correspond to $|6P_{3/2},F=4\rangle$ and
$|6S_{1/2},F=4\rangle$, respectively. The two excited states are assumed to
decay with the rates
$\Gamma_{e}=\Gamma_{b}=\gamma=2\pi\times5.2\,\mathrm{MHz}$. 

\begin{figure}
\includegraphics[width=0.6\textwidth]{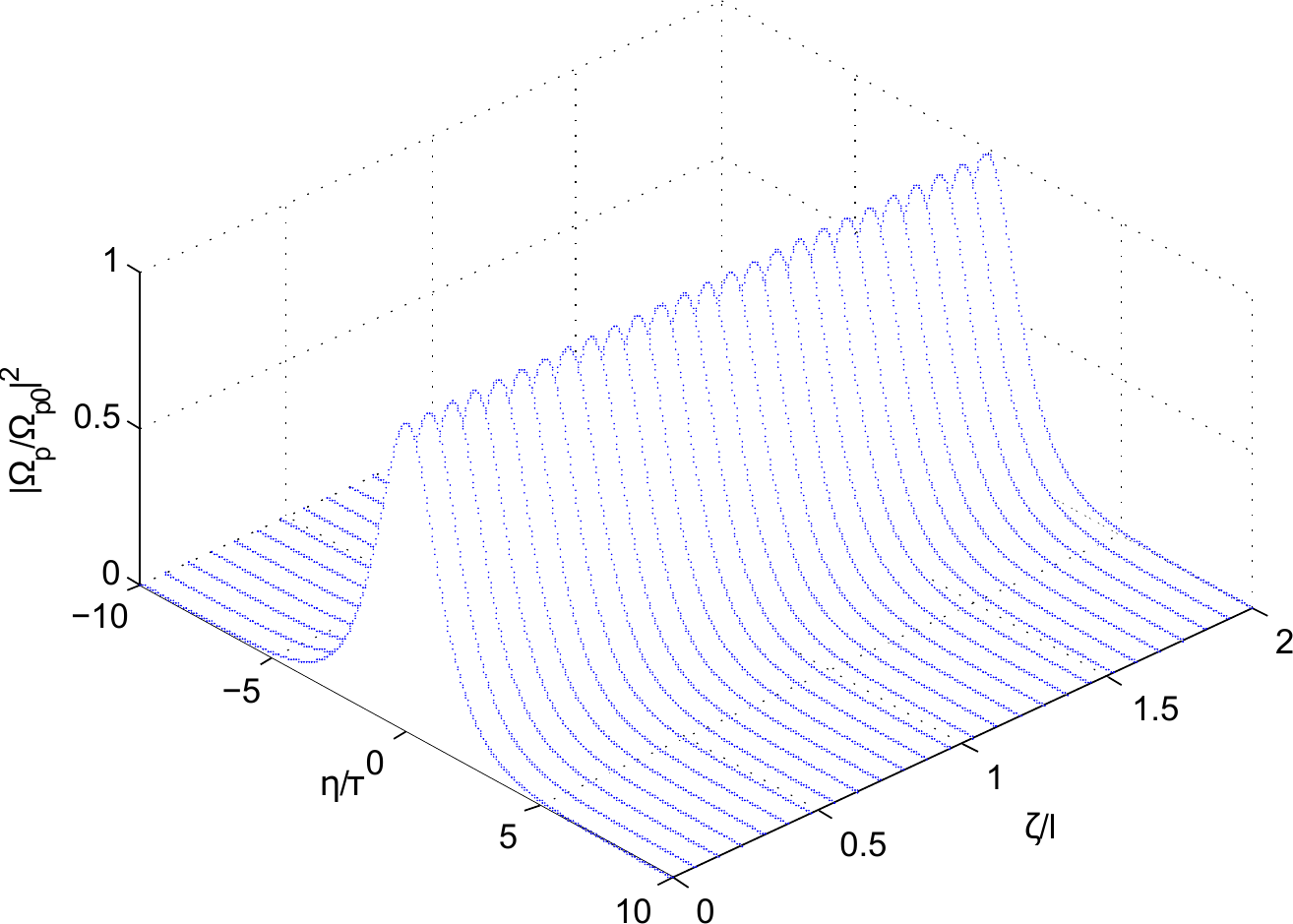}
\caption{Propagation dynamics of an ultraslow optical soliton with
  $\tau=10^{-7}\,\mathrm{s}$, $l=1\,\mathrm{cm}$, and
  $\eta=1.0\times10^{10}\,\mathrm{cm}^{-1}\cdot\mathrm{s}^{-1}$ and the
  parameters given in the main text.}
\label{fig:soliton}
\end{figure}

Assuming the parameteric situation (a) described in the previous section, we
take $|\Omega_{1}|=|\Omega_{2}|=1.97\times10^{9}\,\mathrm{s}^{-1}$,
$|\Omega_{3}|=2.3\times10^{9}\,\mathrm{s}^{-1}$,
$|\Omega_{4}|=16.4\times10^{7}\,\mathrm{s}^{-1}$,
$\Delta_{2}=6.4\times10^{9}\,\mathrm{s}^{-1},$
$\Delta_{p}=5.9\times10^{9}\,\mathrm{s}^{-1}$ and
$\Delta_{3}=82\times10^{7}\,\mathrm{s}^{-1}$. Consequentlhy we obtain
$\kappa_{0}\approx(3.9+0.008i)\,\mathrm{cm}^{-1}$,
$\kappa_{1}\approx(4.7-2.1\times10^{-2}i)10^{-9}\,\mathrm{cm}^{-1}\cdot\mathrm{s}$,
$\kappa_{2}\approx(-8.06-3.6\times10^{-2}i)10^{-17}\,\mathrm{cm}^{-1}\cdot\mathrm{s}^{2}$,
and
$\Theta\approx(3.6-1.12\times10^{-2}i)10^{-19}\,\mathrm{cm}^{-1}\cdot\mathrm{s}^{2}$.
In this case, the standard nonlinear Schrodinger equation (\ref{eq:40-2}) with
$\kappa_{2r}\Theta_{r}<0$ is well characterized, leading to the formation of
dark solitons in the proposed system. With this set of parameters, the
fundamental soliton has a width and amplitude satisfying
$|\Omega_{p0}\tau|=\sqrt{\kappa_{2r}/\Theta_{r}}\simeq4.7$. As shown in
figure~\ref{fig:soliton}, the dark soliton of this type remains fairly stable
during propagation, which is due to the balance between the group-velocity
dispersion and Kerr-type optical nonlinearity.  According to equation~(\ref{eq:71})
in \ref{sec:appendix-C}, the group velocity $v_{g}$ has a general form
$v_{g}^{-1}=c^{-1}+\eta(-g_{1}/Q+S_{1}g_{2}/Q^{2})$, with all coefficients
given in \ref{sec:appendix-C}. With the above system parameters, one
can find $v_{g}\approx7\times10^{-3}c,$ indicating that the soliton propagates
with a slow velocity. 

The formation and propagation of such a slow light optical soliton in the
system is due to the EIT condition described in the situation (a) of the
previous section, i.e., $\beta\neq0$ and $\alpha\neq0$, or equivalently
$\Omega_{1}\Omega_{4}\neq\Omega_{2}\Omega_{3}$. Due to the EIT, the absorption
of the probe field becomes negligibly small.  In this case, an enhanced Kerr
nonlinearity can compensate the dispersion effects in such a highly resonant
medium resulting in shape preserving slow light optical solitons. 

\section{Concluding Remarks\label{sec:concl}}

In conclusion we have demonstrated the existence of dark states which are
essential for appearance of electromagnetically induced transparency (EIT) for
a situation where the atom- light interaction represents a five-level combined
tripod and $\Lambda$ configuration. The EIT is possible in the combined tripod
and $\Lambda$ scheme when the Rabi frequencies of the control fields obey the
condition $\beta\neq0$, $\alpha\neq0$, where $\beta$ and $\alpha$ given by
equations~(\ref{eq:beta})--(\ref{eq:alpha}) characterize the relative amplitudes and
phases of the four control fields. Under this condition, the medium supports
the lossless propagation of slow light. It is analytically demonstrated that
combined tripod and $\Lambda$ scheme can reduce to simpler atom light-coupling
configurations under various quantum interference situations. In particular,
this scheme is equivalent to a four-level $N$-type scheme when $\beta=0$ and
$\alpha\neq0$. On the other hand, for $\beta\neq0$ but $\alpha=0$, a
three-level $\Lambda$-type atom-light coupling scheme can be established.  As a
result, by changing the Rabi frequencies of control fields, it is possible to
make a transition from one limiting case to the another one. This can lead to
switching from subluminality accompanied by EIT to superluminality along with
absorption and visa versa. Based on the coupled Maxwell-Bloch equations, a
nonlinear equation governing the evolution of the probe pulse envelope is then
obtained. This leads to formation of stable optical solitons with a slow
propagating velocity due to the balance between dispersion and Kerr
nonlinearity of the system.

A possible realistic experimental realization of
the proposed combined tripod and $\Lambda$ setup can be implemented for the Cs
atoms. The lower levels $|a\rangle$, $|c\rangle$ and $|d\rangle$ can then be
assigned to $|6S_{1/2},F=4\rangle$, $|6S_{1/2},F=3,M_{F}=+1\rangle$ and
$|6S_{1/2},F=3,M_{F}=+3\rangle$, respectively. Two excited states
$|b\rangle$and $|e\rangle$ can be attributed to the Cs states
$|6P_{3/2},F=4\rangle$ and $|6P_{3/2},F=2,M_{F}=+2\rangle$, respectively. 

\ack

The presented work has been supported by the Lithuanian Research Council
(No.~VP1-3.1-\v{S}MM-01-V-03-001).

\appendix

\section{Eigenstates and eigenvalues for situation (a)\label{sec:appendix-A}}

The expressions for the eigenstates and their corresponding eigenvalues
for situaion (a) are:
\begin{eqnarray}
|n_{1}\rangle =\frac{\alpha^{*}(S-Y)}{\Omega(X-Y)}|b\rangle
-\frac{\sqrt{S-Y}}{\sqrt{2}\Omega}|B_{e}\rangle
-\frac{\alpha^{*}\beta\sqrt{2(S-Y)}}{\Omega(X-Y)}|D_{e}\rangle+|e\rangle\,,\label{eq:43}\\
|n_{2}\rangle =\frac{\alpha^{*}(S-Y)}{\Omega(X-Y)}|b\rangle
+\frac{\sqrt{S-Y}}{\sqrt{2}\Omega}|B_{e}\rangle
+\frac{\alpha^{*}\beta\sqrt{2(S-Y)}}{\Omega(X-Y)}|D_{e}\rangle+|e\rangle\,,\label{eq:44}\\
|n_{3}\rangle =\frac{\alpha^{*}(S+Y)}{\Omega(X+Y)}|b\rangle
-\frac{\sqrt{S+Y}}{\sqrt{2}\Omega}|B_{e}\rangle
-\frac{\alpha^{*}\beta\sqrt{2(S+Y)}}{\Omega(X+Y)}|D_{e}\rangle+|e\rangle\,,\label{eq:45}\\
|n_{4}\rangle =\frac{\alpha^{*}(S+Y)}{\Omega(X+Y)}|b\rangle
+\frac{\sqrt{S+Y}}{\sqrt{2}\Omega}|B_{e}\rangle
+\frac{\alpha^{*}\beta\sqrt{2(S+Y)}}{\Omega(X+Y)}|D_{e}\rangle+|e\rangle\,,\label{eq:46}
\end{eqnarray}
with eigenvalues
\begin{eqnarray}
\lambda_{1} =-\frac{\sqrt{S-Y}}{\sqrt{2}}\,,\label{eq:47}\\
\lambda_{2} =\frac{\sqrt{S-Y}}{\sqrt{2}}\,,\label{eq:48}\\
\lambda_{3} =-\frac{\sqrt{S+Y}}{\sqrt{2}}\,,\label{eq:49}\\
\lambda_{4} =\frac{\sqrt{S+Y}}{\sqrt{2}}\,,\label{eq:50}
\end{eqnarray}
where
\begin{eqnarray}
S =|\alpha|^{2}+|\beta|^{2}+\Omega^{2}\,,\label{eq:51}\\
X =|\alpha|^{2}-|\beta|^{2}+\Omega^{2}\,,\label{eq:52}\\
Y =\sqrt{S^{2}-4|\beta|^{2}\Omega^{2}}\,.\label{eq:53}
\end{eqnarray}

\section{Eigenstates and eigenvalues for situation (b)\label{sec:appendix-B}}

The expressions for the eigenstates and their corresponding eigenvalues
for situaion (b) are:
\begin{eqnarray}
|n_{1}\rangle =-\frac{\Omega}{\alpha}|b\rangle+|e\rangle\,,\label{eq:54}\\
|n_{2}\rangle =|D_{e}\rangle\,,\label{eq:55}\\
|n_{3}\rangle =\frac{\alpha^{*}}{\Omega}|b\rangle-\frac{\sqrt{|\alpha|^{2}
    +\Omega^{2}}}{\Omega}|B_{e}\rangle+|e\rangle\,,\label{eq:56}\\
|n_{4}\rangle =\frac{\alpha^{*}}{\Omega}|b\rangle
+\frac{\sqrt{|\alpha|^{2}+\Omega^{2}}}{\Omega}|B_{e}\rangle+|e\rangle\,,\label{eq:57}
\end{eqnarray}
with eigenvalues
\begin{eqnarray}
\lambda_{1} =0\,,\label{eq:58}\\
\lambda_{2} =0\,,\label{eq:59}\\
\lambda_{3} =-\sqrt{|\alpha|^{2}+\Omega^{2}}\,,\label{eq:60}\\
\lambda_{4} =\sqrt{|\alpha|^{2}+\Omega^{2}}\,.\label{eq:61}
\end{eqnarray}

\section{Explicit expressions for $\kappa_{0}$ , $1/v_{g}$ and
  $\kappa_{2}$\label{sec:appendix-C}}

Expressions for $\kappa_{0}$ , $1/v_{g}$ and $\kappa_{2}$ read
\begin{eqnarray}
\kappa_{0} =\eta\frac{S_{1}}{Q}\,,\label{eq:70}\\
\frac{1}{v_{g}} =\frac{1}{c}+\eta\left(-\frac{g_{1}}{Q}
  +\frac{S_{1}g_{2}}{Q^{2}}\right)\,,\label{eq:71}\\
\kappa_{2} =\eta\left(-\frac{g_{3}}{Q}+\frac{1}{Q^{2}}\left(2g_{1}g_{2}
    -S_{1}g_{4}\right)-\frac{S_{1}g_{2}^{2}}{Q^{3}}\right)\,,\label{eq:72}
\end{eqnarray}
with
\begin{eqnarray}
  \fl g_{1} =t_{2}^{2}+2t_{2}t_{3}-\left(|\Omega_{3}|^{2}
  +|\Omega_{4}|^{2}\right),\label{eq:73}\\
\fl g_{2} =\left(t_{2}+t_{3}\right)\left(|\Omega_{1}|^{2}+|\Omega_{2}|^{2}\right)
+\left(t_{1}+t_{2}\right)\left(|\Omega_{3}|^{2}+|\Omega_{4}|^{2}\right)
-t_{1}t_{2}^{2}-t_{3}t_{2}^{2}-2t_{1}t_{2}t_{3},\label{eq:74}\\
\fl g_{3} =-6\Delta_{2}+2\Delta_{3}+6\Delta_{p}+\rmi\Gamma_{e},\label{eq:75}\\
\fl g_{4} =-2\left(|\Omega_{1}|^{2}+|\Omega_{2}|^{2}+|\Omega_{3}|^{2}
  +|\Omega_{4}|^{2}\right)+2\left(2t_{1}t_{2}+t_{1}t_{3}+t_{2}t_{3}\right)
+2t_{2}^{2},\label{eq:76}
\end{eqnarray}
where $t_{1},t_{2},t_{3}$, $S_{1}$ and $Q$ can be obtained by substituting
$\omega=0$ in coefficients $t_{1}(\omega)$, $t_{2}(\omega)$, $t_{3}(\omega)$,
$S_{1}(\omega)$ and $Q(\omega)$, respectively.

\section{Explicit expressions of $F_{ba}$, $F_{ca}$, $F_{da}$ and
  $F_{ea}$\label{sec:appendix-D}}

After some algebraic calculations, the solutions of
equations~(\ref{eq:22})\textendash (\ref{eq:25}) can be obtained as
\begin{eqnarray}
F_{ba} =\frac{-\Lambda_{p}S_{1}(\omega)}{Q(\omega)},\label{eq:77}\\
F_{ca} =\frac{\Lambda_{p}S_{2}(\omega)}{Q(\omega)},\label{eq:78}\\
F_{da} =\frac{\Lambda_{p}S_{3}(\omega)}{Q(\omega)},\label{eq:79}\\
F_{ea} =\frac{\Lambda_{p}S_{4}(\omega)}{Q(\omega)},\label{eq:80}
\end{eqnarray}
where
\begin{eqnarray}
  \fl S_{1}(\omega)= t_{2}^{2}(\omega)t_{3}(\omega)-t_{2}(\omega)\left(|\Omega_{3}|^{2}
  +|\Omega_{4}|^{2}\right),\label{eq:81}\\
\fl S_{2}(\omega)= |\Omega_{1}||\Omega_{4}|^{2}-|\Omega_{2}||\Omega_{3}||\Omega_{4}|
\rme^{-\rmi\phi}-|\Omega_{1}|t_{2}(\omega)t_{3}(\omega),\label{eq:82}\\
\fl S_{3}(\omega)= |\Omega_{2}||\Omega_{3}|^{2}-|\Omega_{1}||\Omega_{3}||\Omega_{4}|
\rme^{\rmi\phi}-|\Omega_{2}|t_{2}(\omega)t_{3}(\omega),\label{eq:83}\\
\fl S_{4}(\omega)= t_{2}(\omega)\left(|\Omega_{2}||\Omega_{4}|+|\Omega_{1}||\Omega_{3}|
  \rme^{\rmi\phi}\right),\label{eq:84}\\
\fl Q(\omega)= t_{2}(\omega)t_{3}(\omega)\left(|\Omega_{1}|^{2}+|\Omega_{2}|^{2}\right)
+t_{1}(\omega)t_{2}(\omega)\left(|\Omega_{3}|^{2}+|\Omega_{4}|^{2}\right)\nonumber \\
 -t_{1}(\omega)t_{2}^{2}(\omega)t_{3}(\omega)-|\Omega_{2}|^{2}|\Omega_{3}|^{2}
 -|\Omega_{1}|^{2}|\Omega_{4}|^{2}\nonumber\\
 +2|\Omega_{1}||\Omega_{2}||\Omega_{3}||\Omega_{4}|
 \cos(\phi).\label{eq:85}
\end{eqnarray}

\section{Kerr nonlinear coefficient\label{sec:appendix-E}}

One may write the Maxwell equations under the slowly varying envelope
approximation as
\begin{equation}
\frac{\partial\Omega_{p}}{\partial z}+c^{-1}\frac{\partial\Omega_{p}}{\partial t}=
\rmi\eta\Phi_{b}\Phi_{a}^{*},\label{eq:86}
\end{equation}
where $\Phi_{b}\Phi_{a}^{*}=\rho_{ba}$, as well as $\Phi_{a}$ and $\Phi_{b}$
(together with $\Phi_{c}$, $\Phi_{d}$ and $\Phi_{e}$) represent the amplitudes
of atomic wavefunctions for each atomic state and satisfy the relation
\begin{equation}
|\Phi_{a}|^{2}+|\Phi_{b}|^{2}+|\Phi_{c}|^{2}+|\Phi_{d}|^{2}+|\Phi_{e}|^{2}=1.\label{eq:87}
\end{equation}
Initially all atoms are assumed to be in the ground state $|a\rangle$.  As the
Rabi-frequency of the probe field is much weaker than that of the control
fields, one can neglect the depletion of ground level $|a\rangle$, and one has
$\Phi_{a}\thickapprox1$. Adopting a perturbation treatment of the system
response to the first order of probe field, we can take
$\Phi_{L}=\Sigma_{L}\Phi_{L}^{(k)}$ $(L=a,b,c,d,e)$.  Here $\Phi_{L}^{(k)}$ is
the $k$th order part of $\Phi_{L}$ in terms of $\Omega_{p}$, where
$\Phi_{b}^{(0)}=\Phi_{c}^{(0)}=\Phi_{d}^{(0)}=\Phi_{e}^{(0)}=0$, and
$\Phi_{a}^{(0)}=1$, while $\Phi_{a}^{(1)}=0$. Thus, to the first order in
$\Omega_{p}$ we may write 
\begin{eqnarray}
\Phi_{a} =\Phi_{a}^{(0)},\label{eq:88}\\
\Phi_{b}\Phi_{a}^{*} =\Phi_{b}^{(1)}\Phi_{a}^{(0)*}=\Phi_{b}^{(1)}\Phi_{a}^{(0)}=
\Phi_{b}^{(1)}=\rho_{ba}^{(1)},\label{eq:89}\\
\Phi_{c}\Phi_{a}^{*} =\Phi_{c}^{(1)}\Phi_{a}^{(0)*}=\Phi_{c}^{(1)}\Phi_{a}^{(0)}=
\Phi_{c}^{(1)}=\rho_{ca}^{(1)},\label{eq:90}\\
\Phi_{d}\Phi_{a}^{*} =\Phi_{d}^{(1)}\Phi_{a}^{(0)*}=\Phi_{d}^{(1)}\Phi_{a}^{(0)}=
\Phi_{d}^{(1)}=\rho_{da}^{(1)},\label{eq:91}\\
\Phi_{e}\Phi_{a}^{*} =\Phi_{e}^{(1)}\Phi_{a}^{(0)*}=\Phi_{e}^{(1)}\Phi_{a}^{(0)}=
\Phi_{e}^{(1)}=\rho_{ea}^{(1)}.\label{eq:92}
\end{eqnarray}
In this limit equations~(\ref{eq:87}) reduces to
\begin{equation}
|\Phi_{a}^{(0)}|^{2}+|\Phi_{b}^{(1)}|^{2}+|\Phi_{c}^{(1)}|^{2}+|\Phi_{d}^{(1)}|^{2}
+|\Phi_{e}^{(1)}|^{2}=1.\label{eq:93}
\end{equation}
Using equations~(\ref{eq:88})--(\ref{eq:93}), the right hand side of
wave equation~(\ref{eq:86}) can be represented as
\begin{eqnarray}
  \rmi\eta\Phi_{b}\Phi_{a}^{*} =\rmi\eta\Phi_{b}^{(1)}|\Phi_{a}^{(0)}|^{2}\nonumber \\
 =\rmi\eta\Phi_{b}^{(1)}\left[1-(|\Phi_{b}^{(1)}|^{2}+|\Phi_{c}^{(1)}|^{2}
   +|\Phi_{d}^{(1)}|^{2}+|\Phi_{e}^{(1)}|^{2})\right]\nonumber \\
 =\rmi\eta\rho_{ba}^{(1)}-i\eta\rho_{ba}^{(1)}(|\rho_{ba}^{(1)}|^{2}
 +|\rho_{ca}^{(1)}|^{2}+|\rho_{da}^{(1)}|^{2}+|\rho_{ea}^{(1)}|^{2})\,.
 \label{eq:94}
\end{eqnarray}
The first term $\rmi\eta\rho_{ba}^{(1)}$ shows the linear part of the right
hand side of wave equation~(\ref{eq:86}) which was featured in
equation~(\ref{eq:8}). In
addition, $-\rmi\eta\mathrm{Kerr}=-\rmi\eta\rho_{ba}^{(1)}(|\rho_{ba}^{(1)}|^{2}
+|\rho_{ca}^{(1)}|^{2}+|\rho_{da}^{(1)}|^{2}+|\rho_{ea}^{(1)}|^{2})$
represents the nonlinear part of the right hand side of wave
equation~(\ref{eq:86}).  From this expression, the explicit form of the
nonlinear coefficient $\mathrm{Kerr}$ can be readily derived as
\begin{equation}
\mathrm{Kerr}=\rho_{ba}^{(1)}(|\rho_{ba}^{(1)}|^{2}+|\rho_{ca}^{(1)}|^{2}
+|\rho_{da}^{(1)}|^{2}+|\rho_{ea}^{(1)}|^{2}).\label{eq:36-1}
\end{equation}
As $F_{ij}$ is the Fourier transform of $\rho_{ij}^{(1)}$, the coefficients
for $\rho_{ij}^{(1)}$ can be obtained by taking $\omega=0$ in the
coefficients given in \ref{sec:appendix-D}. Replacing the
coefficients for $\rho_{ij}^{(1)}$ in this way into equation~(\ref{eq:36-1})
yields
\begin{equation}
\mathrm{Kerr}=\frac{-S_{1}}{Q|Q|^{2}}\left(|S_{1}|^{2}+|S_{2}|^{2}+|S_{3}|^{2}
  +|S_{4}|^{2}\right)\,.\label{eq:36b-1}
\end{equation}

\section*{References}

\providecommand{\newblock}{}


\begin{thebibliography}{10}
\expandafter\ifx\csname url\endcsname\relax
  \def\url#1{{\tt #1}}\fi
\expandafter\ifx\csname urlprefix\endcsname\relax\def\urlprefix{URL }\fi
\providecommand{\eprint}[2][]{\url{#2}}

\bibitem{Arimondo-96}
Arimondo E 1996 {\em Progress in Optics\/} (Amsterdam: Elsevier)

\bibitem{Harris-PT-1997}
Harris S~E 1997 {\em Phys. Today\/} {\bf 50} 36

\bibitem{Lukin2003}
Lukin M~D 2003 {\em Rev. Mod. Phys.\/} {\bf 75} 457--472

\bibitem{Fleischhauer2005}
Fleischhauer M, Imamoglu A and Marangos J~P 2005 {\em Rev. Mod. Phys.\/} {\bf
  77} 633--673

\bibitem{Wu2005EIT}
Wu Y and Yang X 2005 {\em Phys. Rev. A\/} {\bf 71} 053806

\bibitem{Fleischhauer2016}
Fleischhauer M and Juzeli{\={u}}nas G 2016 Slow, stored and stationary light
  {\em Optics in Our Time\/} ed Al-Amri M~D, El-Gomati M and Zubairy M~S (Cham:
  Springer International Publishing) pp 359--383 ISBN 978-3-319-31903-2

\bibitem{Hau-99}
Hau L~V, Harris S~E, Dutton Z and Behroozi C~H 1999 {\em Nature\/} {\bf 397}
  594

\bibitem{Fleischhauer2000}
Fleischhauer M and Lukin M~D 2000 {\em Phys. Rev. Lett.\/} {\bf 84} 5094--5097

\bibitem{Liu-01}
Liu C, Dutton Z, Behroozi C~H and Hau L~V 2001 {\em Nature\/} {\bf 409} 490

\bibitem{Phill2001}
Phillips D~F, Fleischhauer A, Mair A, Walsworth R~L and Lukin M~D 2001 {\em
  Phys. Rev. Lett.\/} {\bf 86} 783

\bibitem{Fleischhauer2002}
Fleischhauer M and Lukin M~D 2002 {\em Phys. Rev. A\/} {\bf 65} 022314

\bibitem{Juzeliunas-02}
Juzeli{\=u}nas G and Carmichael H~J 2002 {\em Phys. Rev. A\/} {\bf 65} 021601

\bibitem{Bajcsy2003}
Bajcsy M, Zibrov A~S and Lukin M~D 2003 {\em Nature\/} {\bf 426} 638

\bibitem{Yu2009}
Lin Y~W, Liao W~T, Peters T, Chou H~C, Wang J~S, Cho H~W, Kuan P~C and Yu I~A
  2009 {\em Phys. Rev. Lett.\/} {\bf 102} 213601

\bibitem{Wu2003}
Wu Y, Saldana J and Zhu Y 2003 {\em Phys. Rev. A\/} {\bf 67} 013811

\bibitem{Zhang2008}
Zhang Y, Anderson B and Xiao M 2008 {\em Phys. Rev. A\/} {\bf 77} 061801

\bibitem{Zhang2009}
Zhang Y, Khadka U, Anderson B and Xiao M 2009 {\em Phys. Rev. Lett.\/} {\bf
  102} 013601

\bibitem{wu2004}
Wu Y and Deng L 2004 {\em Phys. Rev. Lett.\/} {\bf 93} 143904

\bibitem{Huang2005}
Huang G, Deng L and Payne M~G 2005 {\em Phys. Rev. E\/} {\bf 72} 016617

\bibitem{Li2010}
Li L and Huang G 2010 {\em Phys. Rev. A\/} {\bf 82} 023809

\bibitem{Si2010}
Si L~G, Yang W~X, L\"u X~Y, Hao X and Yang X 2010 {\em Phys. Rev. A\/} {\bf 82}
  013836

\bibitem{Yang2011}
Yang W~X, Chen A~X, Lee R~K and Wu Y 2011 {\em Phys. Rev. A\/} {\bf 84} 013835

\bibitem{Chen2014}
Chen Y, Bai Z and Huang G 2014 {\em Phys. Rev. A\/} {\bf 89} 023835

\bibitem{Joshi2003}
Joshi A, Brown A, Wang H and Xiao M 2003 {\em Phys. Rev. A\/} {\bf 67} 041801

\bibitem{li2006}
Li J~H, L\"u X~Y, Luo J~M and Huang Q~J 2006 {\em Phys. Rev. A\/} {\bf 74}
  035801

\bibitem{Schmidt1996}
Schmidt H and Imamoglu A 1996 {\em Opt. Lett.\/} {\bf 21} 1936--1938

\bibitem{Hai2002}
Wang H, Goorskey D and Xiao M 2002 {\em Opt. Lett.\/} {\bf 27} 258--260

\bibitem{Niu2006}
Niu Y and Gong S 2006 {\em Phys. Rev. A\/} {\bf 73} 053811

\bibitem{Sheng2011}
Sheng J, Yang X, Wu H and Xiao M 2011 {\em Phys. Rev. A\/} {\bf 84} 053820

\bibitem{Hamedi2015}
Hamedi H~R and Juzeli\ifmmode~\bar{u}\else \={u}\fi{}nas G 2015 {\em Phys. Rev.
  A\/} {\bf 91} 053823

\bibitem{Harris1998}
Harris S~E and Yamamoto Y 1998 {\em Phys. Rev. Lett.\/} {\bf 81} 3611--3614

\bibitem{Lukin2000}
Lukin M~D and Imamo\ifmmode~\breve{g}\else \u{g}\fi{}lu A 2000 {\em Phys. Rev.
  Lett.\/} {\bf 84} 1419--1422

\bibitem{Wang2006}
Wang Z~B, Marzlin K~P and Sanders B~C 2006 {\em Phys. Rev. Lett.\/} {\bf 97}
  063901

\bibitem{Shiau2011}
Shiau B~W, Wu M~C, Lin C~C and Chen Y~C 2011 {\em Phys. Rev. Lett.\/} {\bf 106}
  193006

\bibitem{Chen2012}
Chen Y~H, Lee M~J, Hung W, Chen Y~C, Chen Y~F and Yu I~A 2012 {\em Phys. Rev.
  Lett.\/} {\bf 108} 173603

\bibitem{Venkataraman2013}
Vivek~Venkataraman K~S and Gaeta A~L 2013 {\em Nature Photonics\/} {\bf 7}
  138--141

\bibitem{Maxwell2013}
Maxwell D, Szwer D~J, Paredes-Barato D, Busche H, Pritchard J~D, Gauguet A,
  Weatherill K~J, Jones M~P~A and Adams C~S 2013 {\em Phys. Rev. Lett.\/} {\bf
  110} 103001

\bibitem{Baur2014}
Baur S, Tiarks D, Rempe G and D\"urr S 2014 {\em Phys. Rev. Lett.\/} {\bf 112}
  073901

\bibitem{Boller1991}
Boller K~J, Imamo\ifmmode~\breve{g}\else \u{g}\fi{}lu A and Harris S~E 1991
  {\em Phys. Rev. Lett.\/} {\bf 66} 2593--2596

\bibitem{Ruseckas2007}
Ruseckas J, Juzeli\ifmmode~\bar{u}\else \={u}\fi{}nas G, \"Ohberg P and Barnett
  S~M 2007 {\em Phys. Rev. A\/} {\bf 76} 053822

\bibitem{Ruseckas2011}
Ruseckas J, Mekys A and Juzeli\ifmmode~\bar{u}\else \={u}\fi{}nas G 2011 {\em
  Phys. Rev. A\/} {\bf 83} 023812

\bibitem{Paspalakis2002}
Paspalakis E and Knight P~L 2002 {\em Journal of Optics B: Quantum and
  Semiclassical Optics\/} {\bf 4} S372

\bibitem{Schnorrberger-PRL-2009}
Schnorrberger U, Thompson J~D, Trotzky S, Pugatch R, Davidson N, Kuhr S and
  Bloch I 2009 {\em Phys. Rev. Lett.\/} {\bf 103} 033003

\bibitem{Unanyan2010}
Unanyan R~G, Otterbach J, Fleischhauer M, Ruseckas J,
  Kudria\ifmmode~\check{s}\else \v{s}\fi{}ov V and Juzeli\ifmmode~\bar{u}\else
  \={u}\fi{}nas G 2010 {\em Phys. Rev. Lett.\/} {\bf 105} 173603

\bibitem{Ruseckas2013}
Ruseckas J, Kudria\ifmmode~\check{s}\else \v{s}\fi{}ov V~c~v, Yu I~A and
  Juzeli\ifmmode~\bar{u}\else \={u}\fi{}nas G 2013 {\em Phys. Rev. A\/} {\bf
  87} 053840

\bibitem{Lee2014}
Lee M~J, Ruseckas J, Lee C~Y, Kudriasov V, Chang K~F, Cho H~W, Juzeli{\=u}nas G
  and Yu I~A 2014 {\em Nat. Commun.\/} {\bf 5} 5542

\bibitem{Ruseckas2011PRA}
Ruseckas J, Kudria{\v s}ov V, Juzeli{\=u}nas G, Unanyan R~G, Otterbach J and
  Fleischhauer M 2011 {\em Phys. Rev. A\/} {\bf 83} 063811

\bibitem{Bao2011coherent}
Bao Q~Q, Zhang X~H, Gao J~Y, Zhang Y, Cui C~L and Wu J~H 2011 {\em Phys. Rev.
  A\/} {\bf 84} 063812

\bibitem{Raczynski-OC-2006}
Raczynski A, Rzepecka M, Zaremba J and Zielinska-Kaniasty S 2006 {\em Opt.
  Commun.\/} {\bf 260} 73

\bibitem{Raczynski-PRA-2007}
Raczynski A, Zaremba J and Zielinska-Kaniasty S 2007 {\em Phys. Rev. A\/} {\bf
  75} 013810

\bibitem{Mazets2017}
Beck S and Mazets I~E 2017 {\em Phys. Rev. A\/} {\bf 95} 013818

\bibitem{Payne2002}
Payne M~G and Deng L 2002 {\em Phys. Rev. A\/} {\bf 65} 063806

\bibitem{Korsunsky1999}
Korsunsky E~A and Kosachiov D~V 1999 {\em Phys. Rev. A\/} {\bf 60} 4996--5009

\bibitem{Shpaisman2005}
Shpaisman H, Wilson-Gordon A~D and Friedmann H 2005 {\em Phys. Rev. A\/} {\bf
  71} 043812

\bibitem{Fleischhaker2008}
Fleischhaker R and Evers J 2008 {\em Phys. Rev. A\/} {\bf 77} 043805

\bibitem{Rajitha2015}
V R~K, Dey T~N, Evers J and Kiffner M 2015 {\em Phys. Rev. A\/} {\bf 92} 023840

\bibitem{Braje2004}
Braje D~A, Bali\ifmmode~\acute{c}\else \'{c}\fi{} V, Goda S, Yin G~Y and Harris
  S~E 2004 {\em Phys. Rev. Lett.\/} {\bf 93} 183601

\bibitem{Dey2007}
Dey T~N and Agarwal G~S 2007 {\em Phys. Rev. A\/} {\bf 76} 015802

\bibitem{Kang2003}
Kang H and Zhu Y 2003 {\em Phys. Rev. Lett.\/} {\bf 91} 093601

\bibitem{Rodrigo2003}
Rodrigo A~Vicencio M~I~M and Kivshar Y~S 2003 {\em Optics Letters\/} {\bf 28}
  1942--1944

\bibitem{Heebner2002}
E~Heebner J, Boyd R~W and Park Q~H 2002 {\em Phys. Rev. E\/} {\bf 65} 036619

\bibitem{Melloni2003}
Andrea~Melloni F~M and Martinelli M 2003 {\em Optics and Photonics News\/} {\bf
  14} 44--48

\bibitem{Liu2004}
Liu X~J, Jing H and Ge M~L 2004 {\em Phys. Rev. A\/} {\bf 70} 055802

\bibitem{Agrawal2001}
Agrawal G~P 2001 {\em Nonlinear Fiber Optics\/} 3rd ed (Academic, New York)

\bibitem{Hasegawa2003}
Hasegawa A and Matsumoto M 2003 {\em Optical Solitons in Fibers\/} (Berlin:
  Springer)

\bibitem{Xie2006}
Xie X~T, Li W~B and Yang W~X 2006 {\em J. Phys. B\/} {\bf 39} 401

\bibitem{WuPRL2004}
Wu Y and Deng L 2004 {\em Phys. Rev. Lett.\/} {\bf 93} 143904

\bibitem{Burger1999}
Burger S, Bongs K, Dettmer S, Ertmer W, Sengstock K, Sanpera A, Shlyapnikov G~V
  and Lewenstein M 1999 {\em Phys. Rev. Lett.\/} {\bf 83} 5198--5201

\bibitem{Denschlag2000}
Denschlag J, Simsarian J~E, Feder D~L, Clark C~W, Collins L~A, Cubizolles J,
  and W~Hagley L~D, Helmerson K, Reinhardt W~P, Rolston S~L, Schneider B~I and
  Phillips W~D 2000 {\em Science\/} {\bf 287} 97

\bibitem{Huang2002}
Huang G, Szeftel J and Zhu S 2002 {\em Phys. Rev. A\/} {\bf 65} 053605

\bibitem{Kivshar1998}
Kivshar Y~S and Luther-Davies B 1998 {\em Phys. Rep.\/} {\bf 298} 81

\bibitem{YuanYao2007}
Lin Y and Lee R~K 2007 {\em Optics Express\/} {\bf 15} 8781

\bibitem{Xie2007}
Xie X~T, Li W, Li J, Yang W~X, Yuan A and Yang X 2007 {\em Phys. Rev. B\/} {\bf
  75} 184423

\bibitem{YangPRA2009}
Yang W~X, Hou J~M, Lin Y and Lee R~K 2009 {\em Phys. Rev. A\/} {\bf 79} 033825

\bibitem{WuOptLett2004}
Wu Y and Deng L 2004 {\em Optics Letters\/} {\bf 29} 2064

\bibitem{Hang2006}
Hang C, Huang G and Deng L 2006 {\em Phys. Rev. E\/} {\bf 74} 046601

\bibitem{Si2009}
Si L~G, Yang W~X, Liu J~B, Li J and Yang X 2009 {\em Optics Express\/} {\bf 17}
  7771

\bibitem{ChaoHang2010}
Hang C and Huang G 2010 {\em Optics Express\/} {\bf 18} 2952

\bibitem{Chengjie2011}
Zhu C and Huang G 2011 {\em Optics Express\/} {\bf 19} 1963

\bibitem{Liu2010}
Liu J~B, Liu N, Shan C~J, Liu T~K and Huang Y~X 2010 {\em Phys. Rev. E\/} {\bf
  81} 036607

\bibitem{Liang2010}
Li L and Huang G 2010 {\em Phys. Rev. A\/} {\bf 82} 023809

\bibitem{Chen2015}
Chen Y, Chen Z and Huang G 2015 {\em Phys. Rev. A\/} {\bf 91} 023820

\bibitem{Scully-Book-1997}
Scully M~O and Zubairy M~S 1997 {\em Quantum Optics\/} (Cambridge: Cambridge
  University Press)

\bibitem{Harris1999}
Harris S~E and Hau L~V 1999 {\em Phys. Rev. Lett.\/} {\bf 82} 4611--4614

\end{thebibliography}
\end{document}